\documentclass[aps,notitlepage,superscriptaddress]{revtex4-2}

\usepackage[T1]{fontenc}
\usepackage[utf8]{inputenc}

\usepackage{graphicx}

\usepackage{amsmath,amssymb,amsfonts,mathtools}
\usepackage{bm}

\usepackage{siunitx}
\DeclareSIUnit{\pg}{pg}
\DeclareSIUnit{\GHz}{GHz}

\usepackage{enumitem}

\usepackage{booktabs}
\usepackage{multirow}

\usepackage{tikz}
\usetikzlibrary{shapes.geometric,arrows,decorations.markings}

\usepackage{microtype}

\usepackage[normalem]{ulem}
\usepackage{xcolor}

\usepackage[colorlinks=true,citecolor=magenta,linkcolor=blue,urlcolor=blue]{hyperref}
\usepackage[nameinlink,capitalise,noabbrev]{cleveref}

\begin{document}

\title{Vortex MTJs with Opto-Electrical and Spin-Diode Nonlinearities as Multifunctional Neuromorphic Platforms}

\author{Felix Oberbauer}
\affiliation{Institut für Physik, Universität Greifswald, 17489 Greifswald, Germany}
\author{Tristan Joachim Winkel}
\affiliation{Institut für Physik, Universität Greifswald, 17489 Greifswald, Germany}
\author{Clara C Wanjura}
\affiliation{Max Planck Institute for the Science of Light, Staudtstraße 2, 91058 Erlangen, Germany}
\author{Maksim Steblii}
\affiliation{INL - International Iberian Nanotechnology Laboratory, Avenida Mestre José Veiga, s/n, 4715-330 Braga, Portugal}
\author{Jakob Walowski}
\affiliation{Institut für Physik, Universität Greifswald, 17489 Greifswald, Germany}
\author{Tim B\"ohnert}
\affiliation{INL - International Iberian Nanotechnology Laboratory, Avenida Mestre José Veiga, s/n, 4715-330 Braga, Portugal}
\author{Ricardo Ferreira}
\affiliation{INL - International Iberian Nanotechnology Laboratory, Avenida Mestre José Veiga, s/n, 4715-330 Braga, Portugal}
\author{Markus M\"unzenberg}
\email{muenzenbem@uni-greifswald.de}
\affiliation{Institut für Physik, Universität Greifswald, 17489 Greifswald, Germany}
\author{Tahereh Sadat Parvini}
\email{Tahereh.Parvini@wmi.badw.de}
\affiliation{Institut für Physik, Universität Greifswald, 17489 Greifswald, Germany}
\affiliation{Walther-Meißner-Institut, Bayerische Akademie der Wissenschaften, Walther-Meißner-Str.8, 85748 Garching, Germany}


\begin{abstract}
The human brain achieves exceptional energy efficiency by co-locating memory and processing, yet reproducing this principle in hardware remains challenging because many neuromorphic devices require standby power, offer limited programmability, or separate state storage from nonlinear computation. Here we demonstrate a multifunctional spintronic platform based on storage-layer-enabled vortex magnetic tunnel junctions (MTJs) that unifies non-volatile weight storage, optoelectrically driven nonlinear computation, and multilevel readout within a single nanopillar. A thermally programmable FM/AFM storage layer retains analog synaptic weights with zero standby power and enables non-volatile tuning of the vortex gyrotropic resonance over a ${\sim}15$~MHz range. Under optoelectrical operation, combined laser heating and dc bias drive the junction into the bias-enhanced tunnel magneto-Seebeck (bTMS) regime, where the thermoelectric response exhibits a pronounced cubic nonlinearity that provides a compact, hardware-native transfer function for weighted analog computation. The electrical and thermoelectric channels switch at matched coercive fields but with distinct signal amplitudes, yielding an effective four-level readout space. Using crossbar-array simulations parameterized by measured device response maps, we evaluate two neuromorphic operating modes---a bTMS mode (optical input, dc-bias weights) and a frequency-multiplexed spin-diode mode (RF-frequency input, RF-power weights)---and obtain image-classification accuracies of $95.4\%$ and $94.9\%$, respectively, comparable to a matched digital single-layer network with sigmoid activations. Smaller 600~nm devices consistently outperform larger ones, identifying nonlinear-response engineering as a key device-level lever for neuromorphic accuracy. Because bTMS and spin-diode rectification coexist in the same junction, a combined operating regime could enable nonlinear multi-input interactions, including quadratic cross-terms, within a single nanoscale element. These results establish vortex MTJs as a unified nanoscale platform in which programmable magnetism, spin caloritronics, and gyrotropic dynamics converge to implement core primitives for neuromorphic computation.
\end{abstract}

\maketitle

\section{Introduction}

The growing energy and data-movement costs of modern artificial intelligence have intensified interest in neuromorphic computing as a hardware route toward brain-inspired, energy-efficient information processing. By co-locating memory and computation in adaptive physical elements, neuromorphic architectures can mitigate the von Neumann bottleneck that constrains conventional transistor-based systems, particularly for inference workloads requiring massive parallelism and low-latency analog operations \cite{davies2019benchmarks, markovic2020physics, keyes1977physical, aimone2022review, Hayward2024neuromorphic}. Realizing this promise, however, requires nanoscale devices that combine non-volatility, tunability, strong nonlinear response, and low-power operation within a single platform. Spintronic nanoscale oscillators are particularly well suited to this role. Magnetic tunnel junction (MTJ)-based spin-torque nano-oscillators (STNOs), spin Hall nano-oscillators (SHNOs), and related oscillator architectures offer compact footprints, CMOS back-end compatibility, and rich magnetization dynamics that can simultaneously provide computational nonlinearity and memory \cite{grollier2020neuromorphic, zhou2021prospect, vincent2015spin, kurenkov2019artificial, shashank2025ptbi}. Within this family, vortex-based STNOs are especially attractive for neuromorphic operation because their topologically protected vortex core provides robust stability, while the gyrotropic mode yields a smooth yet strongly nonlinear analog transfer function that has already been exploited in reservoir-computing and oscillator-network demonstrations \cite{kammerer2011magnetic, jenkins2021electrical, suess2018topologically, imai2022input, tsunegi2019physical, markovic2019reservoir, riou2019temporal}.

While vortex-based oscillators provide the nonlinear dynamics required for reservoir and analog neuromorphic computing, their deployment in reconfigurable hardware depends critically on persistent, non-volatile control of the oscillator operating point. Martins \emph{et al.} showed that electrically set vortex chirality, retained after removal of the initializing signals, can function as a binary non-volatile synapse when read out through the spin-torque diode effect \cite{Martins2021}. Stebliy \emph{et al.} extended this concept to continuous analog frequency tuning by integrating an antiferromagnet/ferromagnet (IrMn/NiFe) exchange-bias storage layer into the MTJ nanopillar: voltage-pulse Joule heating drives the antiferromagnet above its N\'eel temperature, and subsequent field cooling resets the exchange-bias direction, displacing the vortex-core equilibrium and producing a persistent gyrotropic-frequency shift over a $\sim 15$~MHz range without continuous power dissipation \cite{stebliy2024non}. This thermally assisted scheme provides a continuously tunable non-volatile weight, but it is driven exclusively by electrical stimuli. A complementary, non-contact control channel emerges from spin caloritronics \cite{bauer2012caloritronics, kuschel2019tunnel}: laser-induced thermal gradients across MgO tunnel barriers generate magneto-Seebeck voltages whose sign and magnitude depend on magnetic configuration, as established in CoFeB/MgO/CoFeB junctions \cite{walter2011seebeck} and later enhanced in optimized material systems \cite{boehnke2017large}. When combined with a dc bias, this evolves into the bias-enhanced tunnel magneto-Seebeck (bTMS) regime, in which the thermovoltage develops a strongly nonlinear bias dependence that is well captured by a cubic-like response over the experimentally relevant range \cite{boehnke2015off, oberbauer2025hybrid}. The same vortex MTJ geometry also supports the spin-torque diode effect \cite{tulapurkar2005diode, jenkins2016vortex}, a second nonlinear transduction mode in which radio-frequency excitation near the gyrotropic resonance is rectified into a dc voltage; related rectification-based responses have already been leveraged for neuromorphic and RF signal-processing tasks \cite{leroux2021hardware, ross2023multilayer}. However, the combined use of non-volatile exchange-bias reconfiguration, laser-driven bTMS, and spin-diode rectification in a single vortex MTJ platform has not yet been established experimentally.

Here we use CoFeB/MgO/CoFeB vortex MTJ nanopillars with an integrated IrMn-based storage layer and show that non-volatile magnetic reconfiguration, laser-driven thermoelectric readout (TMS/bTMS), and vortex spin-diode rectification can coexist and be independently harnessed within one device architecture. Building on the previously demonstrated non-volatile frequency-programmable platform, we experimentally investigate laser-driven thermoelectric transport in these storage-layer-enabled vortex STNOs and resolve both the linear TMS regime and the bias-enhanced bTMS regime. We show that the thermovoltage scales linearly with laser power and switches hysteretically between parallel and antiparallel states, while simultaneous dc bias drives a pronounced cubic-like nonlinearity that naturally serves as a compact activation-like transfer function. In addition, the combined electrical and thermoelectric readout yields an effective four-level signal space (matched coercive switching but distinct resistance and thermovoltage levels), and field-resolved thermovoltage spikes associated with Barkhausen-like reversal events indicate a route toward spike-like signalling relevant to event-driven neuromorphic operation. Using the measured device characteristics, we then construct and evaluate a crossbar-array neuromorphic computing scheme in two operating modes implemented on the same hardware platform—bTMS-based and spin-diode-based—achieving classification performance comparable to a digital single-layer network with sigmoid activations, with the spin-diode mode reaching 94.9\%. Finally, we outline a combined operating regime in which bTMS and spin-diode rectification act simultaneously in one device to enable nonlinear multi-input (including quadratic) feature interaction, with operation-equivalent processing-density estimates of $\sim 3\,\mathrm{TFLOPS\,mm^{-2}}$ at current device dimensions and up to $\sim 30{,}000\,\mathrm{TFLOPS\,mm^{-2}}$ under idealized nanometre-scale scaling assumptions (excluding routing and peripheral overheads). These results position vortex MTJs as a uniquely multifunctional neuromorphic primitive integrating non-volatile magnetic programmability, laser-driven spin-caloritronic control, and vortex-dynamical rectification in a single nanoscale element.

\section{Device Features and Opto-electrical Measurement Setup}
\label{sec:device}

\begin{figure*}[t!]
  \centering
  \includegraphics[width=0.95\textwidth]{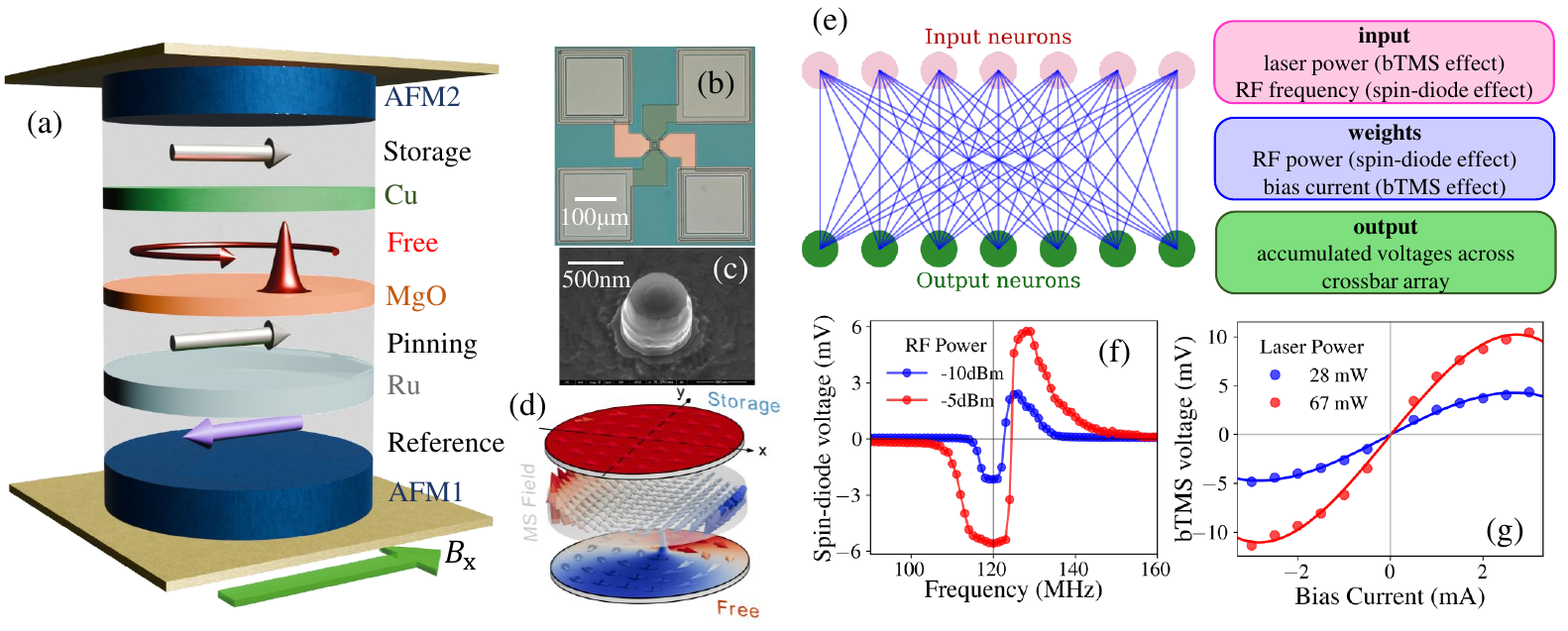}
    \caption{\textbf{Device concept and nonlinear response characteristics enabling neuromorphic functionality.} (a) Schematic representation of the studied structure. (b) Optical microscope image of the structure. (c) Scanning electron microscopy (SEM) image of the MTJ nanopillar. (d) Distribution of the magnetostatic field generated by the storage layer and acting on the free layer, obtained from micromagnetic simulations. (e) Schematic representation of a neural-network layer and the physical quantities that take the roles of input, output, and weights: in the case of the bTMS effect, the input is injected optically while synaptic weights are controlled with bias currents; in the case of the spin-diode effect, the input is encoded in the frequency of an RF signal while its power plays the role of a synaptic weight. The nonlinear device response in both cases enables nonlinear processing, playing a role analogous to a nonlinear activation function. (f) Measured spin-diode response under RF excitation, showing a strongly nonlinear voltage characteristic that directly supplies a rich analog transfer function for neuromorphic signal processing. (g) Measured bias-enhanced tunnel magneto-Seebeck (bTMS) thermovoltage versus dc bias current for two laser powers, highlighting a pronounced cubic-like nonlinearity suitable for neuron-like activation in neuromorphic hardware.}
    \label{Fig1}
\end{figure*}

Vortex spin-transfer torque nano-oscillators (STNOs) were fabricated using magnetic tunnel junction (MTJ) nanopillars with diameters of 600, 800, and 1000~nm, patterned via electron-beam lithography and ion-beam milling. The device platform and spin-diode characterization build on our previously reported non-volatile frequency-programmable vortex MTJ architecture~\cite{stebliy2024non}. The complete stack structure, from bottom to top, consists of [Ta(5)/CuN(25)]$_6$/Ta(5)/Ru(5)/AFM1/SAF/MgO/Free Layer/Cu(10)/Storage Layer/Ta(10)/Ru(7)/TiWN(15)/AlSiCu
(200)/TiWN(15), with all thicknesses in nanometres (Fig.~\ref{Fig1}a--c), deposited onto a 200~mm thermally oxidised silicon wafer [Si/SiO$_2$(200~nm)] by magnetron sputtering using a Singulus TIMARIS Multi-Target-Module system. Sublayers beneath AFM1 were optimized to minimise surface roughness and serve as the bottom electrical contact. Following nanopillar patterning, the structures were embedded in SiO$_2$ for electrical isolation before defining the top contacts. The magnetic stack comprises three functional subsystems. AFM1 [Ir$_{20}$Mn$_{80}$(15), $T_\mathrm{N}^\mathrm{AFM1} \approx 230\,^{\circ}$C] exchange-biases the synthetic antiferromagnet (SAF) reference layer [Co$_{70}$Fe$_{30}$(2)/Ru(0.825)/Co$_{40}$Fe$_{40}$B$_{20}$(2.6)], providing a thermally stable fixed magnetization reference during device operation. The free layer [Co$_{40}$Fe$_{40}$B$_{20}$(2)/Ta(0.21)/NiFe(7)] hosts a magnetic vortex state and is separated from the SAF by an MgO tunnel barrier, yielding a tunnel magnetoresistance ratio of ${\sim}100\%$ and a resistance--area product of ${\sim}10\,\Omega\,\mu\mathrm{m}^2$~\cite{stebliy2024non}. The reconfigurable storage layer [NiFe(6)/Ir$_{20}$Mn$_{80}$(6)] (AFM2, $T_\mathrm{N}^\mathrm{AFM2} \approx 180\,^{\circ}$C) is positioned 10~nm above the free layer via a non-magnetic Cu spacer, generating a tunable magnetostatic stray field that modulates the free-layer vortex dynamics without direct exchange coupling. Both Néel temperatures were determined from continuous-film measurements (with thermal VSM characterization) and are approximate, as patterning may reduce the effective transition temperature. Post-fabrication, the devices were annealed at 330\,$^{\circ}$C for 2~h under an applied in-plane field of $B_x = 1$\,T to establish exchange bias at both FM/AFM interfaces. The \(\sim 50\,^{\circ}\)C separation between $T_\mathrm{N}^\mathrm{AFM1}$ and $T_\mathrm{N}^\mathrm{AFM2}$ defines a selective thermal operating window for non-volatile reconfiguration: voltage pulses (0.9\,V, 1\,ms) generate Joule heating that raises AFM2 above $T_\mathrm{N}^\mathrm{AFM2}$ while leaving AFM1, and therefore the SAF reference, pinned. The storage-layer magnetization then relaxes under an applied reconfiguration field $B_\mathrm{REC}$ and freezes upon cooling, enabling either a saturated single-domain state or a displaced vortex-like configuration with a tunable magnetostatic bias. The resulting storage-layer stray field is tunable over approximately \(\pm 5\)~mT (Fig.~\ref{Fig1}d), and was shown previously to shift the free-layer gyrotropic resonance frequency continuously over a \(\sim 15\)~MHz range in a fully non-volatile manner for 600~nm devices~\cite{stebliy2024non}. The nonlinear device response central to this work is directly evidenced by the spin-torque diode effect, in which an incident RF signal drives gyrotropic precession of the vortex core and the oscillating tunnel magnetoresistance rectifies the response into a DC voltage. Because the rectified amplitude depends nonlinearly on the detuning between the excitation frequency and the vortex resonance, the device naturally implements a compact analog transfer function---analogous to a neuromorphic activation function---without additional circuitry. Fig.~\ref{Fig1}f presents this rectified voltage for a 600\,nm device at nominally zero applied field ($B_\mathrm{ext} \approx 0$) for two RF excitation powers ($-10$\,dBm and $-5$\,dBm), where the pronounced nonlinearity is clearly visible and directly exploited for neuromorphic processing in this work.

In this work, we experimentally heat these devices with a laser and read out the resulting thermoelectric voltages, complementing the non-volatile reconfiguration reported in Ref.~\cite{oberbauer2025magnetic} to realize neuromorphic functionality. For this purpose, a 638~nm diode laser, square-wave modulated at 1~kHz and focused to a $7~\mu$m-diameter spot, periodically heats the MTJ nanopillar (setup detailed in Ref.~\cite{oberbauer2025magnetic}). Unlike pulsed or resonant optical excitation, in which photons couple directly to the magnetic order~\cite{kirilyuk2010ultrafast, winkel2024, parvini2017new, parvini2022, parvini2015, bostrom2021all, vinas2023direct, parvini2025,parvini2026programmable, hamidi2013}, the CW illumination employed here acts exclusively as a thermal actuator via optical absorption and electron--phonon thermalization, leaving the magnetic configuration unperturbed. Within each modulation half-cycle, a quasi-steady thermal profile establishes a steep temperature gradient $\Delta T$ across the MgO barrier. Heat transport through the barrier is dominated by phonons rather than electrons, whose thermal conductance estimated via the Wiedemann--Franz law is orders of magnitude smaller~\cite{sukwon2024thermal,jang2020thermal}. The barrier therefore acts as a thermal bottleneck with effective conductivity $\kappa_{\mathrm{eff}}=\left[\kappa_B^{-1}+(\kappa_I t_{\mathrm{MgO}})^{-1}\right]^{-1}$, where $\kappa_B=4.0~\mathrm{W\,m^{-1}K^{-1}}$ is the thin-film sputtered MgO thermal conductivity---roughly an order of magnitude below single-crystal MgO (${\sim}48~\mathrm{W\,m^{-1}K^{-1}}$) due to grain-boundary and defect scattering---and $\kappa_I=2.5\times10^{7}~\mathrm{W\,m^{-2}K^{-1}}$ is the interface thermal conductance at the CoFeB/MgO boundaries~\cite{teixeira2013giant}. For $t_{\mathrm{MgO}}=2$~nm, this yields $\kappa_{\mathrm{eff}}\approx 0.05~\mathrm{W\,m^{-1}K^{-1}}$, confining virtually the entire temperature drop across the tunnel barrier and minimising unwanted heating of the surrounding metallic layers. The resulting $\Delta T$ asymmetrically broadens the Fermi--Dirac distribution in the heated electrode relative to the cold side. This energy-dependent carrier imbalance, filtered by the MTJ's spin- and energy-selective transmission function $T(E)$, generates a thermoelectric voltage whose magnitude is governed by the logarithmic energy derivative of $T(E)$ at the Fermi level~\cite{walter2011seebeck,czerner2011spin}. Under open-circuit, small-signal conditions, the first-harmonic response obeys \(V_{\mathrm{P/AP}}^{\mathrm{AC}} = S_{\mathrm{P/AP}}\,\Delta T\) and appears at the modulation frequency (TMS).

Applying a dc bias shifts the electrochemical potentials of the electrodes and reshapes the tunneling window, producing a time-averaged offset $V_\mathrm{DC}$ from bias-enhanced TMS (bTMS)~\cite{boehnke2015off,kuschel2019tunnel} while imparting an explicit bias dependence to the AC amplitude through the bias-modified $T(E)$. Within the Onsager linear-response framework~\cite{onsager1931reciprocalI,onsager1931reciprocalII}, the coupled thermal and electrical drives enter as independent thermodynamic forces, and their cross-coupling produces the nonlinear rectification that distinguishes bTMS from pure TMS. Phase-sensitive lock-in detection (1~kHz reference) isolates $V_\mathrm{AC}$, and a parallel dc channel records $V_\mathrm{DC}$. The disappearance of $V_\mathrm{DC}$ at $I_\mathrm{bias}=0$, its sign reversal upon bias inversion, and the fixed phase of $V_\mathrm{AC}$ relative to the optical reference verify a coupled thermal–electrical origin rather than instrumental artifacts. Magnetic-field sweeps show hysteretic switching at identical thresholds in both channels (P$\leftrightarrow$AP), yet the amplitudes differ because the channels weight distinct physics: $V_\mathrm{AC}$ reports the linear Seebeck response set by the spectral slope of $T_m(E)$ near $E_F$, whereas $V_\mathrm{DC}$ reflects nonlinear rectification that requires concurrent thermal and electrical drive. The resulting differences in P:AP ratios provide two orthogonal observables of spin- and energy-resolved transport. In practice, the vector readout $(V_\mathrm{AC},V_\mathrm{DC})$ enables multi-level state discrimination with a single device and, because optical excitation decouples measurement from spin-torque operation, permits independent optimization of thermoelectric contrast and oscillator bias. Together, the TMS (AC) and bTMS (DC) responses constitute a robust, background-resistant platform for reconfigurable spintronic logic and neuromorphic architectures in nanoscale MTJs.

\section{Experimental results and discussion}

The thermoelectric voltage \(V^{\mathrm{AC}}\), arising from the Seebeck effect, is generated by thermally excited charge carriers in response to the temperature difference \(\Delta T\) across the MTJ nanopillars~\cite{uchida2008observation,czerner2011spin,gravier2006thermodynamic}. Its magnitude and sign are governed by the spin-dependent asymmetry of electronic states near the Fermi level, so that the Seebeck coefficient depends on the magnetic configuration of the junction~\cite{walter2011seebeck,schmidt2018boltzmann,boehnke2017large}. As a result, magnetic-field sweeps yield hysteretic switching of \(V^{\mathrm{AC}}\) between the parallel (P) and antiparallel (AP) states, constituting the tunnel magneto-Seebeck (TMS) effect. To probe this behaviour in STNOs, we performed open-circuit magnetic-field sweeps while varying the laser power. The loops show distinct voltages \(V_{\mathrm{P}}^{\mathrm{AC}}\) and \(V_{\mathrm{AP}}^{\mathrm{AC}}\). In each state, the thermovoltage follows the Seebeck relation \(V_{\mathrm{P/AP}}^{\mathrm{AC}} = S_{\mathrm{P/AP}}\,\Delta T\), where \(S_{\mathrm{P}}\) and \(S_{\mathrm{AP}}\) denote the Seebeck coefficients in the P and AP states, respectively. We quantify the thermoelectric contrast using the TMS ratio, defined analogously to the tunnel magnetoresistance (TMR) ratio~\cite{jha2023interface,sadat2023enhancing},
\begin{equation}
\mathrm{TMS} = \frac{S_{\mathrm{AP}} - S_{\mathrm{P}}}{\min\!\left(\lvert S_{\mathrm{AP}}\rvert,\lvert S_{\mathrm{P}}\rvert\right)}
\equiv
\frac{V_{\mathrm{AP}}^{\mathrm{AC}} - V_{\mathrm{P}}^{\mathrm{AC}}}{\min\!\left(\lvert V_{\mathrm{AP}}^{\mathrm{AC}}\rvert,\lvert V_{\mathrm{P}}^{\mathrm{AC}}\rvert\right)}.
\end{equation}

\begin{figure*}[t!]
  \centering
  \includegraphics[width=0.8\textwidth]{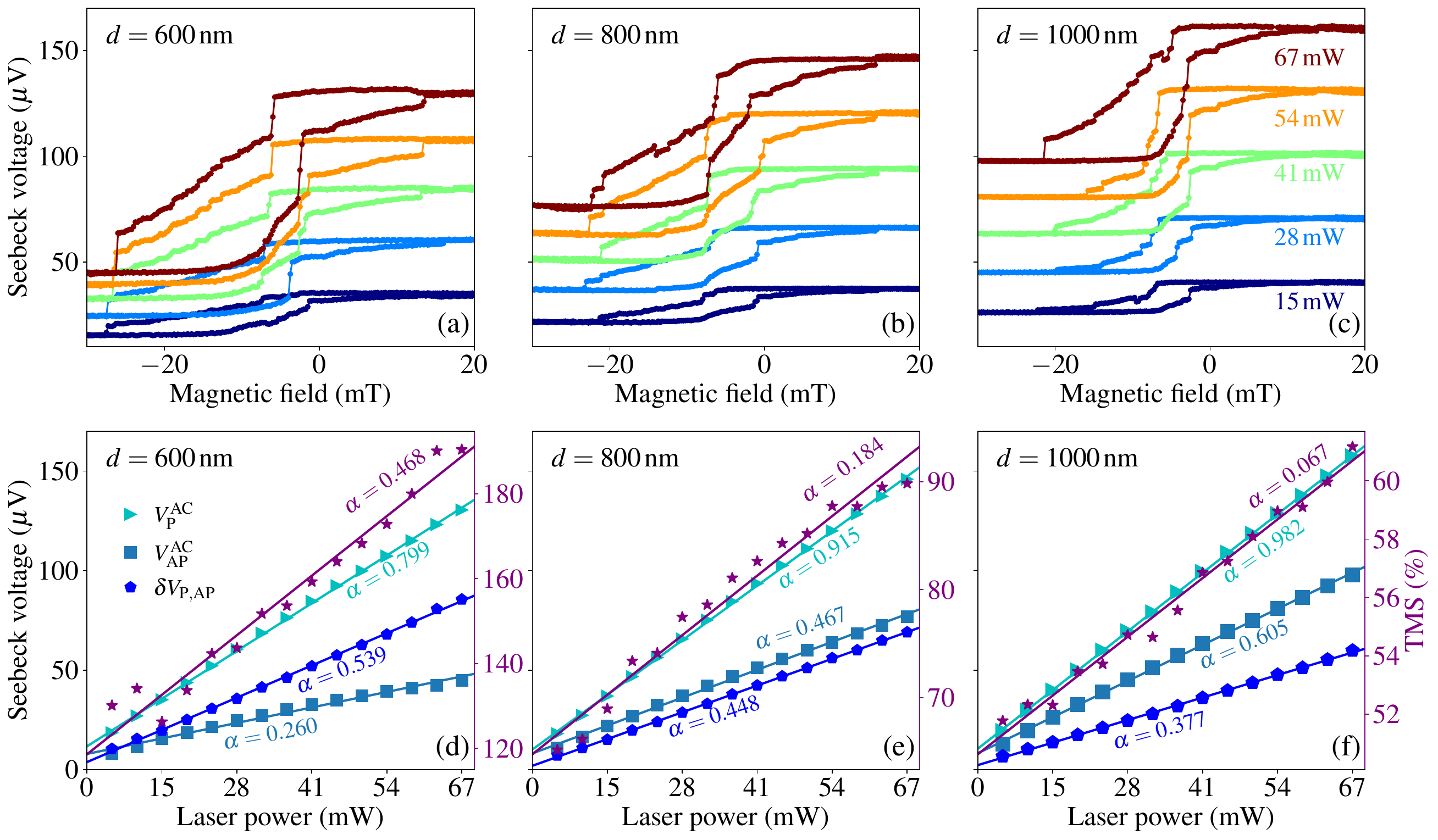}
    \caption{\textbf{Laser-induced thermomagnetic Seebeck effect in vortex spin-torque nano-oscillators.} (a--c) Seebeck voltage \(V^{\mathrm{AC}}\) as a function of in-plane magnetic field \(B_{\parallel}\) for nanopillars with diameters \(d = 600\), 800 and 1000~nm, measured for laser powers at the sample \(P = 15\text{--}67~\mathrm{mW}\). (d--f) Power dependence of the Seebeck voltages in the parallel and antiparallel configurations, \(V_{\mathrm{P}}^{\mathrm{AC}}\) and \(V_{\mathrm{AP}}^{\mathrm{AC}}\), evaluated at \(B_{\parallel}=\pm 30~\mathrm{mT}\), together with their difference \(\delta V_{\mathrm{P,AP}} = V_{\mathrm{AP}}^{\mathrm{AC}} - V_{\mathrm{P}}^{\mathrm{AC}}\) and the TMS ratio. Solid lines are linear fits; the corresponding slopes \(\alpha\) are given in \(\mu\mathrm{V}\,\mathrm{mW}^{-1}\) and \(\%\;\mathrm{mW}^{-1}\).}\label{Fig2}
\end{figure*}

Figure~\ref{Fig2}(a--c) displays \(V^{\mathrm{AC}}(B)\), with switching fields consistent with the TMR characterization. Figure~\ref{Fig2}(d--f) summarizes the power dependence of \(V_{\mathrm{P}}^{\mathrm{AC}}\), \(V_{\mathrm{AP}}^{\mathrm{AC}}\), their voltage contrast \(\delta V_{\mathrm{P,AP}} = V_{\mathrm{AP}}^{\mathrm{AC}} - V_{\mathrm{P}}^{\mathrm{AC}}\), and the TMS ratio. Across all devices, \(V_{\mathrm{P/AP}}^{\mathrm{AC}}\) varies linearly with laser power, consistent with a steady-state temperature rise \(\Delta T \propto P_{\mathrm{abs}}\) and approximately constant \(S_{\mathrm{P/AP}}\) over the modest heating range~\cite{walter2011seebeck,boehnke2017large}. The resulting signals are in the \(\mu\mathrm{V}\) range~\cite{walter2011seebeck,boehnke2017large} and are readily resolved by lock-in detection. Larger-diameter junctions exhibit moderately higher \(\lvert V_{\mathrm{P/AP}}^{\mathrm{AC}}\rvert\), consistent with enhanced optical absorption and heat generation in the illuminated stack~\cite{oberbauer2025magnetic}. Because the response remains linear, the signal can be increased without altering the transport regime; for example, transparent ITO top contacts have been shown to raise thermovoltages into the \(\mathrm{mV}\) range at comparable laser powers by improving optical coupling to the stack~\cite{oberbauer2025magnetic}. More generally, continuous optical tunability provides a practical non-contact analog control parameter relevant for neuromorphic weight programming, reconfigurable spintronic logic, and thermal sensing.

The TMS ratio and the voltage contrast \(\delta V_{\mathrm{P,AP}} = \left(S_{\mathrm{AP}}-S_{\mathrm{P}}\right)\Delta T\) exhibit the opposite diameter dependence, with smaller pillars systematically yielding larger values. This trend reflects the competition between vertical heat flow across the MgO barrier and lateral heat dissipation into the surrounding SiO$_2$ matrix (\(\kappa_{\mathrm{SiO_2}} \approx 1.4~\mathrm{W\,m^{-1}\,K^{-1}}\)). Because SiO$_2$ is a relatively poor thermal conductor, lateral heat spreading is limited; reducing the pillar diameter therefore confines the heating more effectively and increases the vertical temperature gradient across MgO, resulting in a larger \(\Delta T\) at fixed incident power. The same heat-flow picture can also account for the opposite diameter trend reported by B\"ohnert~\emph{et al.}~\cite{boehnke2017Bohnert}, who observed thermovoltage increasing with pillar diameter for junctions embedded in Al$_2$O$_3$ (\(\kappa_{\mathrm{Al_2O_3}} \approx 2~\mathrm{W\,m^{-1}\,K^{-1}}\)). In that case, the higher lateral thermal conductivity enhances sideways heat leakage in small pillars, reducing the effective vertical temperature drop across MgO and allowing wider junctions to sustain a larger fraction of the vertical thermal flux. The sign of the diameter dependence is therefore set by the competition between vertical and lateral heat transport, with the surrounding dielectric determining which pathway dominates. A second contribution to the diameter-dependent TMS ratio is electronic in origin and relates to coherent spin filtering in the MgO barrier. In sputtered polycrystalline MgO, grain boundaries and structural disorder introduce additional, less spin-selective tunneling channels that partially bypass the coherent \(\Delta_1\) symmetry-filtered transport responsible for both high TMR and large Seebeck contrast~\cite{bean2017atomic,mckenna2018stability}. Larger junctions statistically encompass more such defective regions, which can progressively reduce the effective contrast \(S_{\mathrm{AP}}-S_{\mathrm{P}}\). A quantitative separation of thermal and electronic contributions would require finite-element extraction of \(\Delta T(d)\) combined with transport modelling that captures realistic barrier disorder; this is beyond the scope of the present work.

Based on Onsager’s transport framework~\cite{onsager1931reciprocalI,onsager1931reciprocalII}, the lock-in detected laser on/off voltage difference is defined as \(\Delta V_{\mathrm{AC}}^{\mathrm{P/AP}} \equiv V_{\mathrm{AC,on}}^{\mathrm{P/AP}} - V_{\mathrm{AC,off}}^{\mathrm{P/AP}}\) and, under simultaneous heating and bias, obeys
\begin{equation}
\begin{aligned}
\Delta V_{\mathrm{AC}}^{\mathrm{P/AP}}
&=
S_{\mathrm{P/AP}}\,\Delta T
+\left(R_{\mathrm{P/AP}}-\Delta R_{\mathrm{P/AP}}\right) I
-
R_{\mathrm{P/AP}} I,
\label{DeltaV}
\end{aligned}
\end{equation}
where \(R_{\mathrm{P/AP}}\) is the resistance of the unheated STNO and \(\Delta R_{\mathrm{P/AP}}\) denotes the resistance change induced by laser heating in the P and AP states. To quantify the thermoelectric switching contrast, we define the bias-enhanced tunnel magneto-Seebeck (bTMS) ratio as
\begin{equation}
\mathrm{bTMS}
=
\frac{\Delta V_{\mathrm{AC}}^{\mathrm{AP}}-\Delta V_{\mathrm{AC}}^{\mathrm{P}}}
{\min\!\left(\left|\Delta V_{\mathrm{AC}}^{\mathrm{P}}\right|,\ \left|\Delta V_{\mathrm{AC}}^{\mathrm{AP}}\right|\right)}.
\end{equation}

\begin{figure}[t!]
  \centering
  \includegraphics[width=0.5\textwidth]{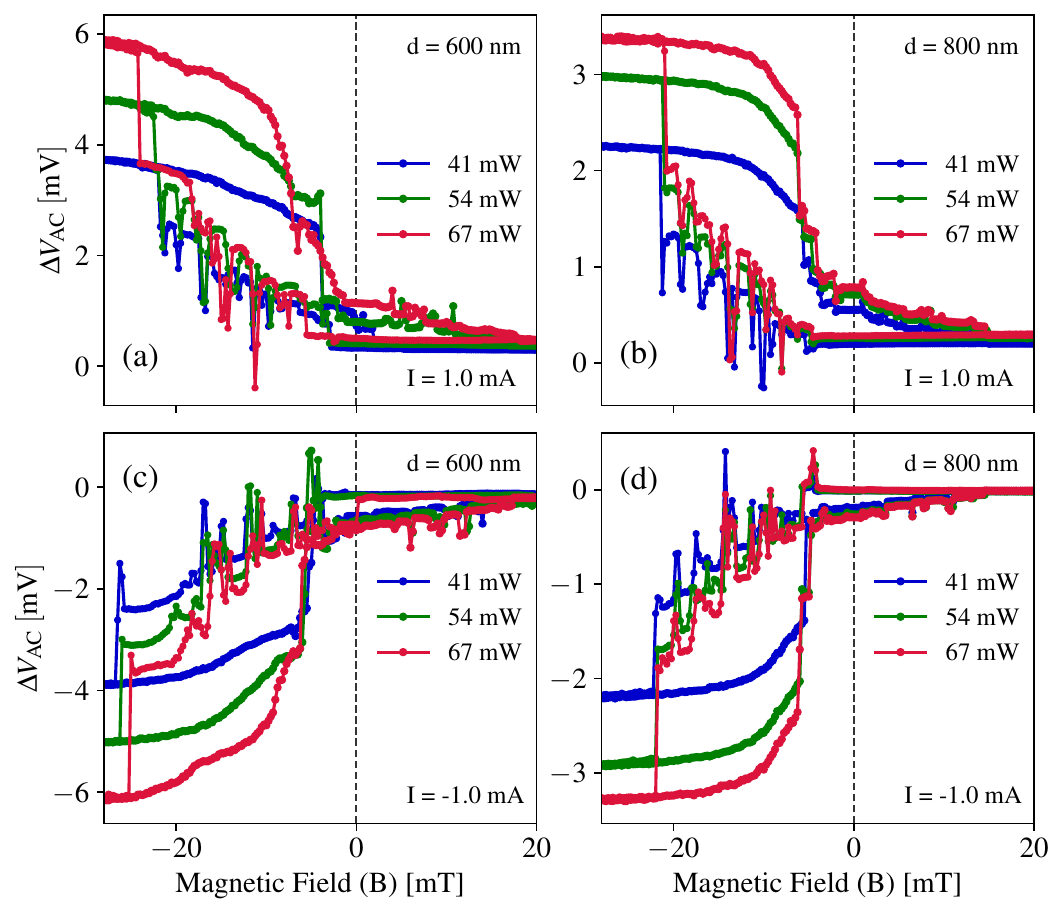}
    \caption{\textbf{Thermovoltage \(\Delta V_{\mathrm{AC}}\) as a function of in-plane magnetic field \(B\).} (a,b) Vortex STNOs with \(d=600\,\mathrm{nm}\) and (c,d) \(d=800\,\mathrm{nm}\). The data were acquired with a magnetic-field step of \(0.25\,\mathrm{mT}\).}
\label{Fig3}
\end{figure}

\begin{figure*}[t!]
  \includegraphics[width=0.90\textwidth]{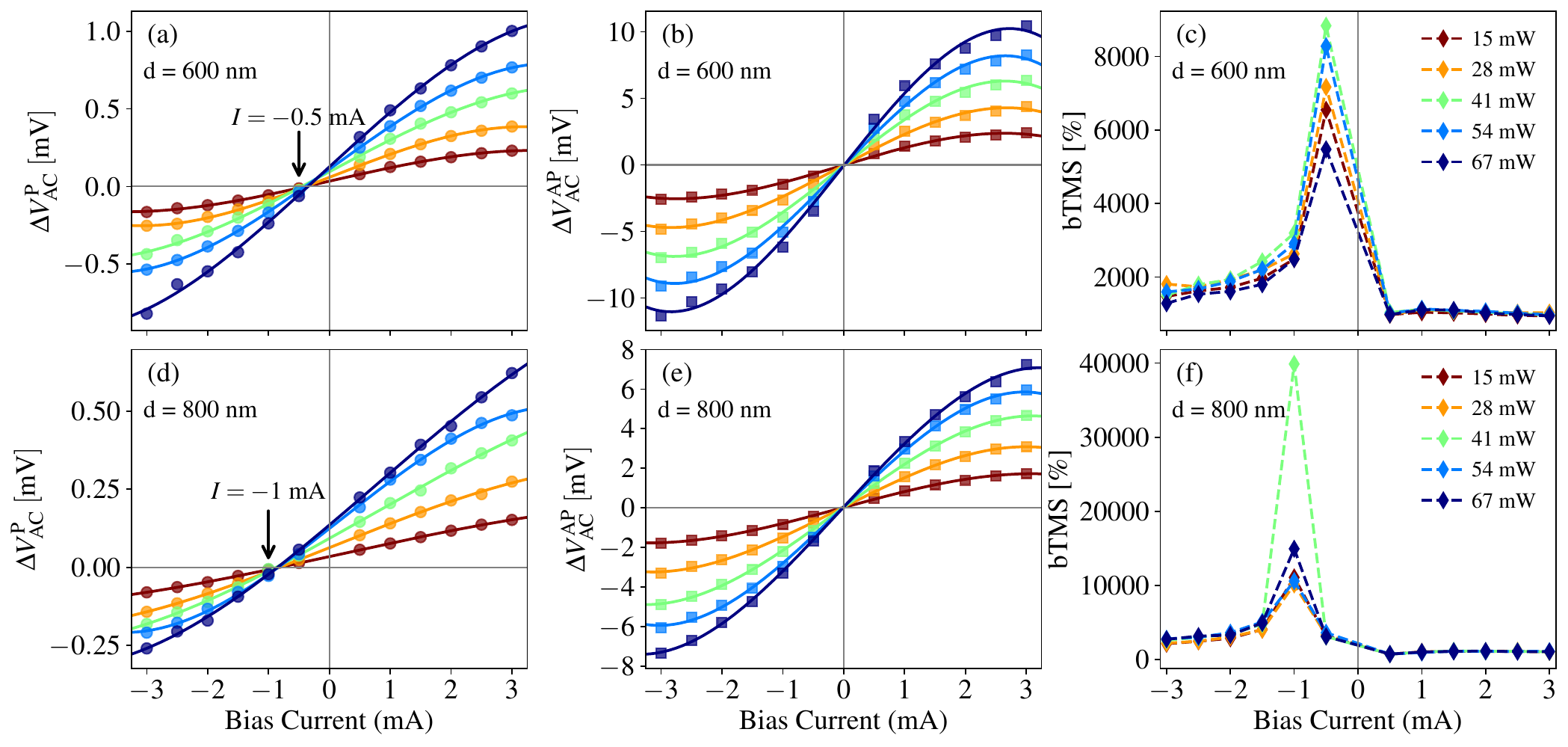}
    \caption{\textbf{Measured thermovoltage in the P and AP states as well as bTMS ratio as a function of bias current for individual laser powers.}  Vortex STNO devices with nanopillar diameters $d = 600$~nm (top row) and $d = 800$~nm (bottom row). Solid lines in (a), (b), (d), and (e) represent cubic polynomial fits to the data of the form $V(I) = \sum_{n=0}^{3} a_n I^n = a_3 I^3 + a_2 I^2 + a_1 I + a_0$, while the curves in (c) and (f) serve as guides to the eye.}
\label{Fig4}
\end{figure*}

As shown in Supplementary Fig.~S5, the resistance and thermoelectric voltage ($\Delta V_{\mathrm{AC}}$) switch at the same magnetic field, confirming their shared magnetic origin. However, the junction remains intrinsically bistable (P/AP), while the combined readout $(R,\Delta V_{\mathrm{AC}})$ provides a two-dimensional signal space that can yield four clearly distinguishable signal clusters in practice. This intrinsic property introduces a multistate encoding mechanism, where information can be simultaneously processed through both charge transport and thermoelectric signals, offering potential advantages for energy-efficient memory and logic architectures. A particularly striking feature is the appearance of Barkhausen jumps in resistance~\cite{kuepferling2015vortex, bohn2018playing}, which manifest as sharp spikes in $\mathrm{V_{\text{AC}}}$, see Fig.~\ref{Fig3}. These spikes are directly correlated with vortex-core transitions~\cite{yoo2012radial}, where rapid domain wall motion during magnetization reversal induces abrupt modifications in spin-dependent transport. As described by Eq.~\ref{DeltaV}, the thermoelectric voltage depends on both the Seebeck effect and resistance variations, with these transient spikes predominantly driven by sudden changes in $\Delta R_{\mathrm{P/AP}}$. The role of the storage layer is also significant, as its magnetostatic stray field alters vortex-core stability and domain wall pinning, further influencing these sharp resistance and thermovoltage fluctuations. Unlike the stable P- and AP-states that serve as conventional readout states in spintronic memory, these transient thermoelectric responses offer a new means of detecting magnetization dynamics in real-time. Their presence suggests potential applications in event-driven neuromorphic computing, where spike-based information encoding mimics biological neurons, as well as in spintronic reservoir computing, where the nonlinear dynamics of magnetization switching contribute to high-dimensional signal processing. Additionally, the ultrafast nature of these thermoelectric transients makes them promising for high-speed magnetic sensing and nonvolatile logic, paving the way for novel functionalities in spin-caloritronic signal processing. Reducing the magnetic field step size enhances the resolution of thermovoltage spikes, making their sharp transitions more apparent. Further improvements in spike characterization could be achieved through time-domain measurements, which may provide insights into the transient nature of pinning-depinning events and spin-texture evolution.

Figure~\ref{Fig4} shows the first-harmonic thermovoltage $\Delta V_{\mathrm{AC}}(I)$ as a function of bias current for several laser powers in the P and AP states, together with the bTMS ratio. Unlike the zero-bias TMS regime, the bTMS thermovoltages are nonlinear in $I$ and are well captured by a cubic polynomial, consistent with the third-order cross-term expected from the Landauer--B\"{u}ttiker framework~\cite{oberbauer2025magnetic, boehnke2015off}. Notably, for $d = 600$~nm ($d = 800$~nm) at $I = -0.5$~mA ($I = -1.0$~mA), the near-vanishing $\Delta V_{\mathrm{AC}}^{\mathrm{P}}$ produces an exceptionally high bTMS ratio, demonstrating that bias tuning can deliver readout contrasts far beyond those accessible by TMR alone~\cite{boehnke2015off}.

\section{Proposal for efficient neuromorphic computing}

\begin{figure*}
	\centering
	\includegraphics[width=\textwidth]{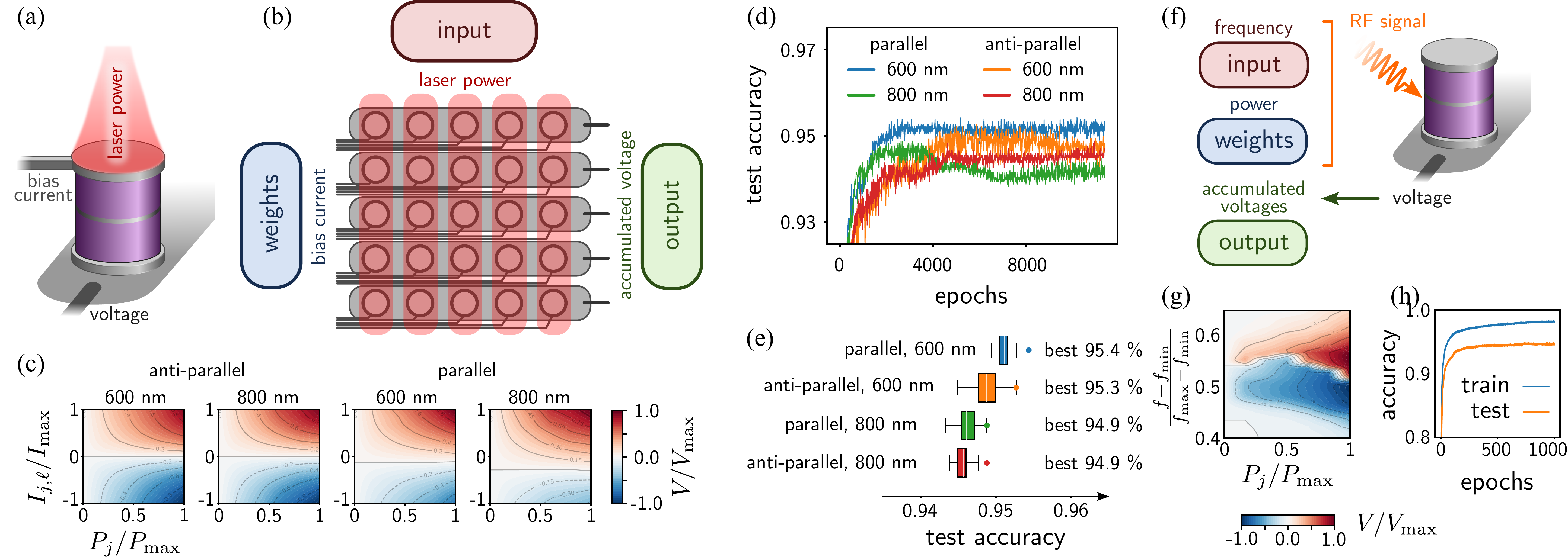}
	\caption{\textbf{Neuromorphic computing performance for different device dimensions and operations.} 
	(a)~We leverage the nonlinear response due to the bTMS effect. The laser power is used to encode the input data, while the bias currents implement synaptic weights.
	(b)~A crossbar-array architecture allows us to perform the linear multiply-accumulate operations as well as the nonlinear activation function in one go. Input powers are set equal across a column, and output voltages are accumulated along a row to yield the output of the neural network layer.
	(c)~We use interpolation functions~\eqref{eq:interpolation} that interpolate linearly between measurement data points to realistically model the device response. We consider different device sizes and operation modes.
	(d)~The evolution of the test accuracy during the training. Smaller devices attain a higher accuracy regardless of the mode of operation. We performed vanilla gradient descent with a learning rate of $10^{-2}$.
	(e)~Best achieved test accuracy (dot) and fluctuations within the $+100$ and $-100$ epochs of the best attained accuracy (box plot).
	(f)~Neuromorphic setup using the spin-diode effect: frequency and power of an incoming RF signal are used to encode the input data (frequency) and the weights (power), while, just as before, the accumulated voltages across a row serve as output.
	(g)~The response of a single device to an RF signal at $B=0$. We only show the response for signal powers $P/P_\mathrm{max}\in[0.4,0.65]$ since outside of this regime, the response is relatively featureless. The frequency axis has been rescaled such that the measurement interval $[f_\mathrm{min},f_\mathrm{max}]$ aligns with $[0,1]$.
	(h)~Training and test accuracy of the simulated neuromorphic chip based on the spin-diode effect.
	}
	\label{fig:neuromorphicProposal}
\end{figure*}%

We recently proposed~\cite{oberbauer2025magnetic} that magnetic tunnel junctions could serve as a flexible and scalable neuromorphic computing platform. Combining the nonlinear device response with a crossbar-array chip architecture would efficiently combine the computation of the multiply-accumulate operations and the nonlinear activation function in one step~\cite{jung2022crossbar}. Together with the small footprint of one tunnel junction, which would allow us to fit millions of devices into one square millimeter, the technology is one of the most scalable approaches currently under investigation. Here, we add another perspective by demonstrating different device operation modes (utilising either the bTMS effect or the spin-diode effect) each inducing a nonlinear device response that can be harnessed for nonlinear processing. Furthermore, we compare different device diameters
and show that our vortex-based spin-torque devices have a more pronounced nonlinear response, which makes them attractive for neuromorphic computing applications.
Finally, we propose that since the bTMS and spin-diode effect are present in the same device, both effects could, in the future, be combined to achieve nonlinear feature processing of multiple inputs---operations that are still typically challenging to achieve with neuromorphic hardware, although it presents an important bottleneck in many standard digital machine learning applications.

\subsection{Neuromorphic computing based on the bTMS effect}
First, we consider the bTMS effect and simulate the operation of a neuromorphic chip based on the measured device response to the laser power and bias current, Fig.~\ref{fig:neuromorphicProposal}~\textbf{a}. The devices are arranged in a crossbar-like configuration, Fig.~\ref{fig:neuromorphicProposal}~\textbf{b}, such that the incident laser powers across the $j$th column are set identical to the $j$th component of the input $x_j = P_{\ell,j} / P_\mathrm{max}$ for all rows $\ell$ with $P_\mathrm{max}$ the maximal input power considered.
The bias currents can be chosen individually for each device and serve as trainable weights $W_{j,\ell} = I_{j,\ell} / I_\mathrm{max}$. Again, the current was rescaled by the maximum considered bias current. The output voltages of each device are collected such that the sum of the voltages across one row can be measured as the accumulated output voltage, which serves as the network output of one neural network layer $y = V / V_\mathrm{max}$. The devices can be operated in a parallel or antiparallel configuration.

To investigate the performance of a neuromorphic chip based on this principle, we simulate the chip operation based on the measured bTMS response and test its performance on a standard classification task of hand-written digit recognition~\cite{Alpaydin1998Optical}.

To train the simulated chip, we define a differentiable interpolation function $f(x_j, W_{j,\ell})$ in the Python library JAX which interpolates the output voltages of one device linearly between measurement data points as a function of input power (representing the input data $x_j$) and bias current (representing the weights $W_{j,\ell}$)
\begin{align}\label{eq:interpolation}
	f_{j,\ell} (x_j, W_{j,\ell}) & = \frac{V_{j,\ell}}{V_\mathrm{max}} \left( \frac{P_j}{P_\mathrm{max}}, \frac{I_{j,\ell}}{I_\mathrm{max}} \right).
\end{align}
The accumulated values $y_\ell = \sum_j f_{j,\ell} (x_j, W_{j,\ell})$ serve as the $\ell$th output.
We show the interpolation functions for different device diameters and device operations (antiparallel vs. parallel) in Fig.~\ref{fig:neuromorphicProposal}~\textbf{c}.

For our simulations, we consider an array of $64\times10$ magnetic tunnel junctions representing a single neural network layer mapping from the pixel values of the input images to the output vector in which each vector entry represents one class with the maximal entry indicating the class. During training, we minimize the categorical cross-entropy loss function. 
We train the simulated chip by performing differentiation with JAX on the loss function. In the future, we envisage that the training would be performed directly on the chip with physics-based training methods~\cite{momeni2024training} such as Equilibrium propagation~\cite{scellier2017equilibrium} or a method recently proposed for nonlinear systems~\cite{cin2025training}. Such efficient physics-based training methods have the advantage that they allow us to perform the training directly in the hardware, enabling successful training even in the presence of device imperfections.

We train for $12,000$ epochs to ensure that the loss function has converged for all tested configurations. The classification accuracy on the test set during the training is shown in Fig.~\ref{fig:neuromorphicProposal}~\textbf{d}.
We see that the smaller devices at $600\,\mathrm{nm}$ generally perform better than the larger devices at $800\,\mathrm{nm}$. We attribute this to the stronger nonlinear response present in small devices. Furthermore, the maximal test accuracy is slightly higher for devices operated in a parallel configuration than in an antiparallel configuration. This may be due to the smaller training gradients obtained in the antiparallel configuration, such that the training may get stuck in local minima, although such problems could, in principle, be resolved through the use of optimizers such as Adam. The smaller training gradients also lead to a slower initial increase of the test accuracy. 
We show the best attained test accuracies in Fig.~\ref{fig:neuromorphicProposal}~\textbf{e} as well as the fluctuations over the $200$ epochs around the best attained accuracy. 
Overall, the test accuracies are much higher than in the previously investigated devices~\cite{oberbauer2025magnetic} and are now comparable to the accuracy of $95\,\%$ which we achieved by a single-layer digital artificial neural network of the same size with sigmoid activations. We attribute this improvement to the more pronounced nonlinear response in the present devices. Hence, the nonlinear device response is a resource for neuromorphic computing.

\subsection{Neuromorphic computing based on the spin-diode effect}
Next, we consider the spin-diode effect for the purpose of neuromorphic computing by utilising the device's nonlinear response to an RF signal. Concretely, we encode the input data in the RF frequency while we encode the weights in the power, Fig.~\ref{fig:neuromorphicProposal}~\textbf{f}. As before, we consider a crossbar arrangement of the devices similar to Fig.~\ref{fig:neuromorphicProposal}~\textbf{b}. The accumulated voltages across one row determine one component of the output vector. This encoding has the advantage that, in contrast to the previous setup, which required bias currents to encode the weights, here, both input and weights are encoded in the RF signal, which implies that devices can potentially have an even smaller footprint since it is not necessary to attach cables to each individual signal. Furthermore, changing the RF power has a lower latency than changing the bias currents, which can potentially result in a speed-up of an in-situ training process.

Again, we use a differentiable interpolation function based on the measurements to simulate the device response shown in Fig.~\ref{fig:neuromorphicProposal}~\textbf{g} for a single device and train the simulated neuromorphic chip on the image classification task. During training, we compute the gradients w.r.t. the signal power of each device. We initialise the powers in the range $P/P_\mathrm{max}\in[0.47,0.57]$ since outside of this range, the response is more featureless, which would result in small gradients and which we empirically found hinders and in some cases even prevents the training.
We show the training and test accuracy of the neuromorphic system during the training in Fig.~\ref{fig:neuromorphicProposal}~\textbf{h} with a best attained test accuracy of $94.9\%$. 
This demonstrates that the spin-diode effect is yet another important resource for neuromorphic computing, with the added advantage that chips based on the spin-diode effect have a potentially lower spatial footprint and lower latency than devices based on the bTMS effect. On the other hand, generating the individual RF signals for each device requires complex hardware, such as vector network analysers with multiple ports, which limits the scalability of this approach and leads to bulky hardware. In contrast, the bTMS effect is based on a thermovoltage induced with a compact laser array directly above the device, which can be implemented and addressed efficiently.

\subsection{Combining the bTMS and the spin-diode effect}
Remarkably, both the bTMS effect and the spin-diode effect produce voltage changes of similar magnitude.
Since the bTMS effect and the spin-diode effect are present in the same device, we propose combining them in the future. A single device would then be subject to pulsed optical heating, a bias current, and an RF signal, such that the induced voltage in the device will depend nonlinearly on four parameters which can be reconfigured during the device operation: the laser power, bias current, the RF power, and RF frequency. Hence, the device would be able to implement a nonlinear function
\begin{equation}
    g(x_1,x_2,W_1,W_2) = \frac{V}{V_\mathrm{max}}\left(
        \frac{P^\mathrm{laser}}{P^\mathrm{laser}_\mathrm{max}},
        \frac{f-f_\mathrm{min}}{f_\mathrm{max}-f_\mathrm{min}},
        \frac{I}{I_\mathrm{max}},
        \frac{P^\mathrm{RF}}{P^\mathrm{RF}_\mathrm{max}}
    \right).
\end{equation}
In general, the function $g$ is expected to be highly nonlinear in $x_1$ and $x_2$. To motivate the power of such an approach, we consider the Taylor expansion of $g$ around suitable $x_1^{(0)}$, $x_2^{(0)}$ which is of the form $g(x_1,x_2,W_1,W_2) = g_0 + g_1^{(1)} x_1 + g_1^{(2)} x_2 + g_2 x_1 x_2 + \mathcal{O}(x_1^2) + \mathcal{O}(x_2^2)$, so in a suitable regime this approach could already enable quadratic processing, combining multiple inputs within a single device.
Many modern machine learning operations indeed rely on the nonlinear interaction between features (inputs), such as the attention mechanism, recurrent neural networks, or graph neural networks~\cite{vaswani2017attention, hochreiter1997lstm, kipf2017gcn}.

Such operations are typically computationally costly to perform on standard digital computers and are among the computational bottlenecks in conventional digital computing. Performing these operations in a single device, or a crossbar array of devices, could significantly accelerate these operations in the future.
Considering furthermore the small device footprint (which can be optimized down to a few nanometers per device) and operations at MHz to GHz rates, this would enable a high computational density. Concretely, we can estimate the number of floating point operations required to match the multiplication and summation steps that could be performed by a crossbar array combining the bTMS and spin-diode effect: Along one row of the array consisting of $N$ devices, the neuromorphic chip would be able to perform the equivalent of $3N-1$ floating point operations ($2N$ multiplications and $N-1$ summations). At the current fabricated footprint of around $25\,(\mu\mathrm{m})^2$ per device, a crossbar array yields around $40{,}000$ devices per mm$^2$, corresponding to a processing density of $120\,\mathrm{GFLOPS\,mm^{-2}}$ at MHz operation rates. With an industry-realistic array footprint of around $1\,\mu\mathrm{m}^2$ per device, and accounting for inter-device spacing, this rises to roughly ${\sim}3\,\mathrm{TFLOPS\,mm^{-2}}$. In comparison, state-of-the-art silicon digital processors currently achieve compute densities in the range of $10$--$50\,\mathrm{GFLOPS\,mm^{-2}}$. Overall, this presents MTJs as an attractive platform for neuromorphic computing, to be explored in future research, especially concerning processing operations that involve the nonlinear interaction between features.

\section{Conclusion}

We have established a storage-layer-enabled vortex MTJ as a multifunctional neuromorphic device that unifies, within a single nanopillar architecture, three capabilities hitherto treated in isolation: non-volatile magnetic programmability, laser-driven spin-caloritronic operation, and nonlinear vortex-dynamical rectification. In our previous work~\cite{stebliy2024non}, Joule-heating-driven exchange-bias reconfiguration of the integrated FM/AFM storage layer was shown to enable non-volatile, continuous tuning of the gyrotropic resonance frequency over a $\sim$15~MHz range, thereby implementing a programmable synaptic weight via spin-torque diode rectification whose magnitude is set by the detuning between the RF excitation and the junction resonance. The strongly nonlinear frequency-dependent rectification response of the spin-diode effect provides an intrinsic analog transfer function, establishing the device as a viable neuromorphic primitive. Here, we extend this platform to laser-induced thermal gradients and resolve both the linear tunnel magneto-Seebeck (TMS) and bias-enhanced bTMS regimes within the same nanopillar geometry. In the linear TMS regime, the thermovoltage scales linearly with laser power and remains readily detectable by standard CMOS-compatible electronics. In the bTMS regime, the simultaneous application of optical heating and dc bias drives a pronounced cubic nonlinearity in the thermoelectric response, constituting a compact analog activation function in hardware. Concurrently, the electrical and thermoelectric readout channels switch synchronously at matched coercive fields but with distinct amplitudes, producing an effective four-level signal space within a single nominally two-state junction. Barkhausen-related thermovoltage spikes during magnetization reversal further demonstrate that the thermoelectric channel is sensitive to abrupt vortex-core and domain-wall dynamics, providing a natural spike-based signaling pathway directly relevant to event-driven neuromorphic operation.

Using the measured device characteristics, we evaluate neuromorphic computing in two operating modes: a bTMS-based mode, in which optical power encodes the input and dc bias encodes synaptic weights, and a spin-diode-based mode, in which RF frequency encodes the input and RF power encodes the weights. In both cases, computation relies on intrinsic device nonlinearities within a crossbar-style analog architecture. Both modes achieve classification accuracies comparable to a digital single-layer network with sigmoid activations, with the best bTMS configuration reaching $95.4\%$ and the spin-diode mode reaching $94.9\%$. The consistently stronger performance of smaller-diameter devices confirms that engineering the nonlinear response is a key device-level design lever for neuromorphic accuracy. Because bTMS and spin-diode rectification coexist in the same vortex MTJ, we further propose a combined operating regime in which optical, dc, and RF controls are applied simultaneously to generate nonlinear interactions between multiple inputs---including quadratic cross-terms---within a single nanoscale element. Conservative processing-density estimates yield ${\sim}0.12\,\mathrm{TFLOPS\,mm^{-2}}$ at the current fabricated device footprint of ${\sim}25\,\mu\mathrm{m}^2$, rising to ${\sim}3\,\mathrm{TFLOPS\,mm^{-2}}$ at an industry-realistic array footprint of ${\sim}1\,\mu\mathrm{m}^2$---two to three orders of magnitude above the $10$--$50\,\mathrm{GFLOPS\,mm^{-2}}$ typical of state-of-the-art silicon processors. Together, these results move vortex MTJs beyond single-function oscillators or memory elements, establishing them as a unified nanoscale platform in which programmable magnetism, spin caloritronics, and gyrotropic dynamics converge to implement nonlinear primitives for next-generation neuromorphic hardware.

\begin{acknowledgments}
This project was supported by funding from the European Union's Horizon 2020 research and innovation program under grant agreement No. 899559 (SpinAge). The authors declare no competing interests.
\end{acknowledgments}

\section*{Data Availability}
Data are available from the corresponding author upon reasonable request.

\bibliographystyle{apsrev4-2}
\bibliography{main}

\begin{thebibliography}{65}%
\makeatletter
\providecommand \@ifxundefined [1]{%
 \@ifx{#1\undefined}
}%
\providecommand \@ifnum [1]{%
 \ifnum #1\expandafter \@firstoftwo
 \else \expandafter \@secondoftwo
 \fi
}%
\providecommand \@ifx [1]{%
 \ifx #1\expandafter \@firstoftwo
 \else \expandafter \@secondoftwo
 \fi
}%
\providecommand \natexlab [1]{#1}%
\providecommand \enquote  [1]{``#1''}%
\providecommand \bibnamefont  [1]{#1}%
\providecommand \bibfnamefont [1]{#1}%
\providecommand \citenamefont [1]{#1}%
\providecommand \href@noop [0]{\@secondoftwo}%
\providecommand \href [0]{\begingroup \@sanitize@url \@href}%
\providecommand \@href[1]{\@@startlink{#1}\@@href}%
\providecommand \@@href[1]{\endgroup#1\@@endlink}%
\providecommand \@sanitize@url [0]{\catcode `\\12\catcode `\$12\catcode `\&12\catcode `\#12\catcode `\^12\catcode `\_12\catcode `\%12\relax}%
\providecommand \@@startlink[1]{}%
\providecommand \@@endlink[0]{}%
\providecommand \url  [0]{\begingroup\@sanitize@url \@url }%
\providecommand \@url [1]{\endgroup\@href {#1}{\urlprefix }}%
\providecommand \urlprefix  [0]{URL }%
\providecommand \Eprint [0]{\href }%
\providecommand \doibase [0]{https://doi.org/}%
\providecommand \selectlanguage [0]{\@gobble}%
\providecommand \bibinfo  [0]{\@secondoftwo}%
\providecommand \bibfield  [0]{\@secondoftwo}%
\providecommand \translation [1]{[#1]}%
\providecommand \BibitemOpen [0]{}%
\providecommand \bibitemStop [0]{}%
\providecommand \bibitemNoStop [0]{.\EOS\space}%
\providecommand \EOS [0]{\spacefactor3000\relax}%
\providecommand \BibitemShut  [1]{\csname bibitem#1\endcsname}%
\let\auto@bib@innerbib\@empty
\bibitem [{\citenamefont {Davies}(2019)}]{davies2019benchmarks}%
  \BibitemOpen
  \bibfield  {author} {\bibinfo {author} {\bibfnamefont {M.}~\bibnamefont {Davies}},\ }\href {https://doi.org/10.1038/s42256-019-0097-1} {\bibfield  {journal} {\bibinfo  {journal} {Nat. Mach. Intell}\ }\textbf {\bibinfo {volume} {1}},\ \bibinfo {pages} {386} (\bibinfo {year} {2019})}\BibitemShut {NoStop}%
\bibitem [{\citenamefont {Markovi{\'c}}\ \emph {et~al.}(2020)\citenamefont {Markovi{\'c}}, \citenamefont {Mizrahi}, \citenamefont {Querlioz},\ and\ \citenamefont {Grollier}}]{markovic2020physics}%
  \BibitemOpen
  \bibfield  {author} {\bibinfo {author} {\bibfnamefont {D.}~\bibnamefont {Markovi{\'c}}}, \bibinfo {author} {\bibfnamefont {A.}~\bibnamefont {Mizrahi}}, \bibinfo {author} {\bibfnamefont {D.}~\bibnamefont {Querlioz}},\ and\ \bibinfo {author} {\bibfnamefont {J.}~\bibnamefont {Grollier}},\ }\href {https://doi.org/10.1038/s42254-020-0208-2} {\bibfield  {journal} {\bibinfo  {journal} {Nat. Rev. Phys}\ }\textbf {\bibinfo {volume} {2}},\ \bibinfo {pages} {499} (\bibinfo {year} {2020})}\BibitemShut {NoStop}%
\bibitem [{\citenamefont {Keyes}(1977)}]{keyes1977physical}%
  \BibitemOpen
  \bibfield  {author} {\bibinfo {author} {\bibfnamefont {R.~W.}\ \bibnamefont {Keyes}},\ }\href {https://doi.org/10.1126/science.195.4283.1230} {\bibfield  {journal} {\bibinfo  {journal} {Science}\ }\textbf {\bibinfo {volume} {195}},\ \bibinfo {pages} {1230} (\bibinfo {year} {1977})}\BibitemShut {NoStop}%
\bibitem [{\citenamefont {Aimone}\ \emph {et~al.}(2022)\citenamefont {Aimone}, \citenamefont {Date}, \citenamefont {Fonseca-Guerra}, \citenamefont {Hamilton}, \citenamefont {Henke}, \citenamefont {Kay}, \citenamefont {Kenyon}, \citenamefont {Kulkarni}, \citenamefont {Mniszewski}, \citenamefont {Parsa} \emph {et~al.}}]{aimone2022review}%
  \BibitemOpen
  \bibfield  {author} {\bibinfo {author} {\bibfnamefont {J.~B.}\ \bibnamefont {Aimone}}, \bibinfo {author} {\bibfnamefont {P.}~\bibnamefont {Date}}, \bibinfo {author} {\bibfnamefont {G.~A.}\ \bibnamefont {Fonseca-Guerra}}, \bibinfo {author} {\bibfnamefont {K.~E.}\ \bibnamefont {Hamilton}}, \bibinfo {author} {\bibfnamefont {K.}~\bibnamefont {Henke}}, \bibinfo {author} {\bibfnamefont {B.}~\bibnamefont {Kay}}, \bibinfo {author} {\bibfnamefont {G.~T.}\ \bibnamefont {Kenyon}}, \bibinfo {author} {\bibfnamefont {S.~R.}\ \bibnamefont {Kulkarni}}, \bibinfo {author} {\bibfnamefont {S.~M.}\ \bibnamefont {Mniszewski}}, \bibinfo {author} {\bibfnamefont {M.}~\bibnamefont {Parsa}}, \emph {et~al.},\ }\href {https://doi.org/10.1088/2634-4386/ac889c} {\bibfield  {journal} {\bibinfo  {journal} {Neuromorphic Computing and Engineering}\ }\textbf {\bibinfo {volume} {2}},\ \bibinfo {pages} {032003} (\bibinfo {year} {2022})}\BibitemShut {NoStop}%
\bibitem [{\citenamefont {Hayward}\ \emph {et~al.}(2024)\citenamefont {Hayward} \emph {et~al.}}]{Hayward2024neuromorphic}%
  \BibitemOpen
  \bibfield  {author} {\bibinfo {author} {\bibfnamefont {T.~J.}\ \bibnamefont {Hayward}} \emph {et~al.},\ }\href {https://doi.org/10.1038/s44306-024-00019-2} {\bibfield  {journal} {\bibinfo  {journal} {npj Spintronics}\ }\textbf {\bibinfo {volume} {2}},\ \bibinfo {pages} {19} (\bibinfo {year} {2024})}\BibitemShut {NoStop}%
\bibitem [{\citenamefont {Grollier}\ \emph {et~al.}(2020)\citenamefont {Grollier}, \citenamefont {Querlioz}, \citenamefont {Camsari}, \citenamefont {Everschor-Sitte}, \citenamefont {Fukami},\ and\ \citenamefont {Stiles}}]{grollier2020neuromorphic}%
  \BibitemOpen
  \bibfield  {author} {\bibinfo {author} {\bibfnamefont {J.}~\bibnamefont {Grollier}}, \bibinfo {author} {\bibfnamefont {D.}~\bibnamefont {Querlioz}}, \bibinfo {author} {\bibfnamefont {K.}~\bibnamefont {Camsari}}, \bibinfo {author} {\bibfnamefont {K.}~\bibnamefont {Everschor-Sitte}}, \bibinfo {author} {\bibfnamefont {S.}~\bibnamefont {Fukami}},\ and\ \bibinfo {author} {\bibfnamefont {M.~D.}\ \bibnamefont {Stiles}},\ }\href {https://doi.org/10.1038/s41928-019-0360-9} {\bibfield  {journal} {\bibinfo  {journal} {Nat. Electron.}\ }\textbf {\bibinfo {volume} {3}},\ \bibinfo {pages} {360} (\bibinfo {year} {2020})}\BibitemShut {NoStop}%
\bibitem [{\citenamefont {Zhou}\ and\ \citenamefont {Chen}(2021)}]{zhou2021prospect}%
  \BibitemOpen
  \bibfield  {author} {\bibinfo {author} {\bibfnamefont {J.}~\bibnamefont {Zhou}}\ and\ \bibinfo {author} {\bibfnamefont {J.}~\bibnamefont {Chen}},\ }\href {https://doi.org/10.1002/aelm.202100465} {\bibfield  {journal} {\bibinfo  {journal} {Adv. Electron. Mater.}\ }\textbf {\bibinfo {volume} {7}},\ \bibinfo {pages} {2100465} (\bibinfo {year} {2021})}\BibitemShut {NoStop}%
\bibitem [{\citenamefont {Vincent}\ \emph {et~al.}(2015)\citenamefont {Vincent}, \citenamefont {Larroque}, \citenamefont {Locatelli}, \citenamefont {Romdhane}, \citenamefont {Bichler}, \citenamefont {Gamrat}, \citenamefont {Zhao}, \citenamefont {Klein}, \citenamefont {Galdin-Retailleau},\ and\ \citenamefont {Querlioz}}]{vincent2015spin}%
  \BibitemOpen
  \bibfield  {author} {\bibinfo {author} {\bibfnamefont {A.~F.}\ \bibnamefont {Vincent}}, \bibinfo {author} {\bibfnamefont {J.}~\bibnamefont {Larroque}}, \bibinfo {author} {\bibfnamefont {N.}~\bibnamefont {Locatelli}}, \bibinfo {author} {\bibfnamefont {N.~B.}\ \bibnamefont {Romdhane}}, \bibinfo {author} {\bibfnamefont {O.}~\bibnamefont {Bichler}}, \bibinfo {author} {\bibfnamefont {C.}~\bibnamefont {Gamrat}}, \bibinfo {author} {\bibfnamefont {W.~S.}\ \bibnamefont {Zhao}}, \bibinfo {author} {\bibfnamefont {J.-O.}\ \bibnamefont {Klein}}, \bibinfo {author} {\bibfnamefont {S.}~\bibnamefont {Galdin-Retailleau}},\ and\ \bibinfo {author} {\bibfnamefont {D.}~\bibnamefont {Querlioz}},\ }\href {https://doi.org/10.1109/TBCAS.2015.2414423} {\bibfield  {journal} {\bibinfo  {journal} {IEEE transactions on biomedical circuits and systems}\ }\textbf {\bibinfo {volume} {9}},\ \bibinfo {pages} {166} (\bibinfo {year} {2015})}\BibitemShut {NoStop}%
\bibitem [{\citenamefont {Kurenkov}\ \emph {et~al.}(2019)\citenamefont {Kurenkov}, \citenamefont {DuttaGupta}, \citenamefont {Zhang}, \citenamefont {Fukami}, \citenamefont {Horio},\ and\ \citenamefont {Ohno}}]{kurenkov2019artificial}%
  \BibitemOpen
  \bibfield  {author} {\bibinfo {author} {\bibfnamefont {A.}~\bibnamefont {Kurenkov}}, \bibinfo {author} {\bibfnamefont {S.}~\bibnamefont {DuttaGupta}}, \bibinfo {author} {\bibfnamefont {C.}~\bibnamefont {Zhang}}, \bibinfo {author} {\bibfnamefont {S.}~\bibnamefont {Fukami}}, \bibinfo {author} {\bibfnamefont {Y.}~\bibnamefont {Horio}},\ and\ \bibinfo {author} {\bibfnamefont {H.}~\bibnamefont {Ohno}},\ }\href {https://doi.org/10.1002/adma.201900636} {\bibfield  {journal} {\bibinfo  {journal} {Adv. Mater.}\ }\textbf {\bibinfo {volume} {31}},\ \bibinfo {pages} {1900636} (\bibinfo {year} {2019})}\BibitemShut {NoStop}%
\bibitem [{\citenamefont {Shashank}\ \emph {et~al.}(2025)\citenamefont {Shashank}, \citenamefont {Kumar}, \citenamefont {Parvini}, \citenamefont {Heyen}, \citenamefont {Zeng}, \citenamefont {Yankovich}, \citenamefont {Rajabali}, \citenamefont {Olsson}, \citenamefont {Münzenberg},\ and\ \citenamefont {Åkerman}}]{shashank2025ptbi}%
  \BibitemOpen
  \bibfield  {author} {\bibinfo {author} {\bibfnamefont {U.}~\bibnamefont {Shashank}}, \bibinfo {author} {\bibfnamefont {A.}~\bibnamefont {Kumar}}, \bibinfo {author} {\bibfnamefont {T.~S.}\ \bibnamefont {Parvini}}, \bibinfo {author} {\bibfnamefont {H.}~\bibnamefont {Heyen}}, \bibinfo {author} {\bibfnamefont {L.}~\bibnamefont {Zeng}}, \bibinfo {author} {\bibfnamefont {A.~B.}\ \bibnamefont {Yankovich}}, \bibinfo {author} {\bibfnamefont {M.}~\bibnamefont {Rajabali}}, \bibinfo {author} {\bibfnamefont {E.}~\bibnamefont {Olsson}}, \bibinfo {author} {\bibfnamefont {M.}~\bibnamefont {Münzenberg}},\ and\ \bibinfo {author} {\bibfnamefont {J.}~\bibnamefont {Åkerman}},\ }\href {https://arxiv.org/abs/2507.10219} {\bibinfo {title} {Bulk spin-orbit torque-driven spin hall nano-oscillators using ptbi alloys}} (\bibinfo {year} {2025}),\ \Eprint {https://arxiv.org/abs/2507.10219} {arXiv:2507.10219 [cond-mat.mes-hall]} \BibitemShut {NoStop}%
\bibitem [{\citenamefont {Kammerer}\ \emph {et~al.}(2011)\citenamefont {Kammerer}, \citenamefont {Weigand}, \citenamefont {Curcic}, \citenamefont {Noske}, \citenamefont {Sproll}, \citenamefont {Vansteenkiste}, \citenamefont {Van~Waeyenberge}, \citenamefont {Stoll}, \citenamefont {Woltersdorf}, \citenamefont {Back} \emph {et~al.}}]{kammerer2011magnetic}%
  \BibitemOpen
  \bibfield  {author} {\bibinfo {author} {\bibfnamefont {M.}~\bibnamefont {Kammerer}}, \bibinfo {author} {\bibfnamefont {M.}~\bibnamefont {Weigand}}, \bibinfo {author} {\bibfnamefont {M.}~\bibnamefont {Curcic}}, \bibinfo {author} {\bibfnamefont {M.}~\bibnamefont {Noske}}, \bibinfo {author} {\bibfnamefont {M.}~\bibnamefont {Sproll}}, \bibinfo {author} {\bibfnamefont {A.}~\bibnamefont {Vansteenkiste}}, \bibinfo {author} {\bibfnamefont {B.}~\bibnamefont {Van~Waeyenberge}}, \bibinfo {author} {\bibfnamefont {H.}~\bibnamefont {Stoll}}, \bibinfo {author} {\bibfnamefont {G.}~\bibnamefont {Woltersdorf}}, \bibinfo {author} {\bibfnamefont {C.~H.}\ \bibnamefont {Back}}, \emph {et~al.},\ }\href {https://doi.org/10.1038/ncomms1277} {\bibfield  {journal} {\bibinfo  {journal} {Nat. Commun.}\ }\textbf {\bibinfo {volume} {2}},\ \bibinfo {pages} {279} (\bibinfo {year} {2011})}\BibitemShut {NoStop}%
\bibitem [{\citenamefont {Jenkins}\ \emph {et~al.}(2021)\citenamefont {Jenkins}, \citenamefont {Alvarez}, \citenamefont {Memshawy}, \citenamefont {Bortolotti}, \citenamefont {Cros}, \citenamefont {Freitas},\ and\ \citenamefont {Ferreira}}]{jenkins2021electrical}%
  \BibitemOpen
  \bibfield  {author} {\bibinfo {author} {\bibfnamefont {A.~S.}\ \bibnamefont {Jenkins}}, \bibinfo {author} {\bibfnamefont {L.~S.~E.}\ \bibnamefont {Alvarez}}, \bibinfo {author} {\bibfnamefont {S.}~\bibnamefont {Memshawy}}, \bibinfo {author} {\bibfnamefont {P.}~\bibnamefont {Bortolotti}}, \bibinfo {author} {\bibfnamefont {V.}~\bibnamefont {Cros}}, \bibinfo {author} {\bibfnamefont {P.~P.}\ \bibnamefont {Freitas}},\ and\ \bibinfo {author} {\bibfnamefont {R.}~\bibnamefont {Ferreira}},\ }\href {https://doi.org/10.1038/s42005-021-00614-3} {\bibfield  {journal} {\bibinfo  {journal} {Commun. Phys.}\ }\textbf {\bibinfo {volume} {4}},\ \bibinfo {pages} {107} (\bibinfo {year} {2021})}\BibitemShut {NoStop}%
\bibitem [{\citenamefont {Suess}\ \emph {et~al.}(2018)\citenamefont {Suess}, \citenamefont {Bachleitner-Hofmann}, \citenamefont {Satz}, \citenamefont {Weitensfelder}, \citenamefont {Vogler}, \citenamefont {Bruckner}, \citenamefont {Abert}, \citenamefont {Pr{\"u}gl}, \citenamefont {Zimmer}, \citenamefont {Huber} \emph {et~al.}}]{suess2018topologically}%
  \BibitemOpen
  \bibfield  {author} {\bibinfo {author} {\bibfnamefont {D.}~\bibnamefont {Suess}}, \bibinfo {author} {\bibfnamefont {A.}~\bibnamefont {Bachleitner-Hofmann}}, \bibinfo {author} {\bibfnamefont {A.}~\bibnamefont {Satz}}, \bibinfo {author} {\bibfnamefont {H.}~\bibnamefont {Weitensfelder}}, \bibinfo {author} {\bibfnamefont {C.}~\bibnamefont {Vogler}}, \bibinfo {author} {\bibfnamefont {F.}~\bibnamefont {Bruckner}}, \bibinfo {author} {\bibfnamefont {C.}~\bibnamefont {Abert}}, \bibinfo {author} {\bibfnamefont {K.}~\bibnamefont {Pr{\"u}gl}}, \bibinfo {author} {\bibfnamefont {J.}~\bibnamefont {Zimmer}}, \bibinfo {author} {\bibfnamefont {C.}~\bibnamefont {Huber}}, \emph {et~al.},\ }\href {https://doi.org/10.1038/s41928-018-0084-2} {\bibfield  {journal} {\bibinfo  {journal} {Nat. Electron.}\ }\textbf {\bibinfo {volume} {1}},\ \bibinfo {pages} {362} (\bibinfo {year} {2018})}\BibitemShut {NoStop}%
\bibitem [{\citenamefont {Imai}\ \emph {et~al.}(2022)\citenamefont {Imai}, \citenamefont {Nakajima}, \citenamefont {Tsunegi},\ and\ \citenamefont {Taniguchi}}]{imai2022input}%
  \BibitemOpen
  \bibfield  {author} {\bibinfo {author} {\bibfnamefont {Y.}~\bibnamefont {Imai}}, \bibinfo {author} {\bibfnamefont {K.}~\bibnamefont {Nakajima}}, \bibinfo {author} {\bibfnamefont {S.}~\bibnamefont {Tsunegi}},\ and\ \bibinfo {author} {\bibfnamefont {T.}~\bibnamefont {Taniguchi}},\ }\href {https://doi.org/10.1038/s41598-022-26018-z} {\bibfield  {journal} {\bibinfo  {journal} {Sci. Rep.}\ }\textbf {\bibinfo {volume} {12}},\ \bibinfo {pages} {21651} (\bibinfo {year} {2022})}\BibitemShut {NoStop}%
\bibitem [{\citenamefont {Tsunegi}\ \emph {et~al.}(2019)\citenamefont {Tsunegi}, \citenamefont {Taniguchi}, \citenamefont {Nakajima}, \citenamefont {Miwa}, \citenamefont {Yakushiji}, \citenamefont {Fukushima}, \citenamefont {Yuasa},\ and\ \citenamefont {Kubota}}]{tsunegi2019physical}%
  \BibitemOpen
  \bibfield  {author} {\bibinfo {author} {\bibfnamefont {S.}~\bibnamefont {Tsunegi}}, \bibinfo {author} {\bibfnamefont {T.}~\bibnamefont {Taniguchi}}, \bibinfo {author} {\bibfnamefont {K.}~\bibnamefont {Nakajima}}, \bibinfo {author} {\bibfnamefont {S.}~\bibnamefont {Miwa}}, \bibinfo {author} {\bibfnamefont {K.}~\bibnamefont {Yakushiji}}, \bibinfo {author} {\bibfnamefont {A.}~\bibnamefont {Fukushima}}, \bibinfo {author} {\bibfnamefont {S.}~\bibnamefont {Yuasa}},\ and\ \bibinfo {author} {\bibfnamefont {H.}~\bibnamefont {Kubota}},\ }\href {https://doi.org/10.1063/1.5081797} {\bibfield  {journal} {\bibinfo  {journal} {Appl. Phys. Lett.}\ }\textbf {\bibinfo {volume} {114}},\ \bibinfo {pages} {164101} (\bibinfo {year} {2019})}\BibitemShut {NoStop}%
\bibitem [{\citenamefont {Markovi\'{c}}\ \emph {et~al.}(2019)\citenamefont {Markovi\'{c}} \emph {et~al.}}]{markovic2019reservoir}%
  \BibitemOpen
  \bibfield  {author} {\bibinfo {author} {\bibfnamefont {D.}~\bibnamefont {Markovi\'{c}}} \emph {et~al.},\ }\href {https://doi.org/10.1063/1.5079305} {\bibfield  {journal} {\bibinfo  {journal} {Appl. Phys. Lett.}\ }\textbf {\bibinfo {volume} {114}},\ \bibinfo {pages} {012409} (\bibinfo {year} {2019})}\BibitemShut {NoStop}%
\bibitem [{\citenamefont {Riou}\ \emph {et~al.}(2019)\citenamefont {Riou}, \citenamefont {Torrejon}, \citenamefont {Garitaine}, \citenamefont {Araujo}, \citenamefont {Bortolotti}, \citenamefont {Cros}, \citenamefont {Tsunegi}, \citenamefont {Yakushiji}, \citenamefont {Fukushima}, \citenamefont {Kubota}, \citenamefont {Yuasa}, \citenamefont {Querlioz}, \citenamefont {Stiles},\ and\ \citenamefont {Grollier}}]{riou2019temporal}%
  \BibitemOpen
  \bibfield  {author} {\bibinfo {author} {\bibfnamefont {M.}~\bibnamefont {Riou}}, \bibinfo {author} {\bibfnamefont {J.}~\bibnamefont {Torrejon}}, \bibinfo {author} {\bibfnamefont {B.}~\bibnamefont {Garitaine}}, \bibinfo {author} {\bibfnamefont {F.~A.}\ \bibnamefont {Araujo}}, \bibinfo {author} {\bibfnamefont {P.}~\bibnamefont {Bortolotti}}, \bibinfo {author} {\bibfnamefont {V.}~\bibnamefont {Cros}}, \bibinfo {author} {\bibfnamefont {S.}~\bibnamefont {Tsunegi}}, \bibinfo {author} {\bibfnamefont {K.}~\bibnamefont {Yakushiji}}, \bibinfo {author} {\bibfnamefont {A.}~\bibnamefont {Fukushima}}, \bibinfo {author} {\bibfnamefont {H.}~\bibnamefont {Kubota}}, \bibinfo {author} {\bibfnamefont {S.}~\bibnamefont {Yuasa}}, \bibinfo {author} {\bibfnamefont {D.}~\bibnamefont {Querlioz}}, \bibinfo {author} {\bibfnamefont {M.~D.}\ \bibnamefont {Stiles}},\ and\ \bibinfo {author} {\bibfnamefont {J.}~\bibnamefont {Grollier}},\ }\href {https://doi.org/10.1103/PhysRevApplied.12.024049} {\bibfield  {journal} {\bibinfo  {journal}
  {Phys. Rev. Appl.}\ }\textbf {\bibinfo {volume} {12}},\ \bibinfo {pages} {024049} (\bibinfo {year} {2019})}\BibitemShut {NoStop}%
\bibitem [{\citenamefont {Martins}\ \emph {et~al.}(2021)\citenamefont {Martins}, \citenamefont {Jenkins}, \citenamefont {Alvarez}, \citenamefont {Borme}, \citenamefont {B\"ohnert}, \citenamefont {Ventura}, \citenamefont {Freitas},\ and\ \citenamefont {Ferreira}}]{Martins2021}%
  \BibitemOpen
  \bibfield  {author} {\bibinfo {author} {\bibfnamefont {L.}~\bibnamefont {Martins}}, \bibinfo {author} {\bibfnamefont {A.~S.}\ \bibnamefont {Jenkins}}, \bibinfo {author} {\bibfnamefont {L.~S.~E.}\ \bibnamefont {Alvarez}}, \bibinfo {author} {\bibfnamefont {J.}~\bibnamefont {Borme}}, \bibinfo {author} {\bibfnamefont {T.}~\bibnamefont {B\"ohnert}}, \bibinfo {author} {\bibfnamefont {J.}~\bibnamefont {Ventura}}, \bibinfo {author} {\bibfnamefont {P.~P.}\ \bibnamefont {Freitas}},\ and\ \bibinfo {author} {\bibfnamefont {R.}~\bibnamefont {Ferreira}},\ }\href {https://doi.org/10.1038/s41598-021-95569-4} {\bibfield  {journal} {\bibinfo  {journal} {Sci. Rep.}\ }\textbf {\bibinfo {volume} {11}},\ \bibinfo {pages} {16094} (\bibinfo {year} {2021})}\BibitemShut {NoStop}%
\bibitem [{\citenamefont {Stebliy}\ \emph {et~al.}(2024)\citenamefont {Stebliy}, \citenamefont {Jenkins}, \citenamefont {Benetti}, \citenamefont {Paz},\ and\ \citenamefont {Ferreira}}]{stebliy2024non}%
  \BibitemOpen
  \bibfield  {author} {\bibinfo {author} {\bibfnamefont {M.}~\bibnamefont {Stebliy}}, \bibinfo {author} {\bibfnamefont {A.}~\bibnamefont {Jenkins}}, \bibinfo {author} {\bibfnamefont {L.}~\bibnamefont {Benetti}}, \bibinfo {author} {\bibfnamefont {E.}~\bibnamefont {Paz}},\ and\ \bibinfo {author} {\bibfnamefont {R.}~\bibnamefont {Ferreira}},\ }\href {https://doi.org/10.1002/adfm.202405776} {\bibfield  {journal} {\bibinfo  {journal} {Adv. Funct. Mater.}\ }\textbf {\bibinfo {volume} {34}},\ \bibinfo {pages} {2405776} (\bibinfo {year} {2024})}\BibitemShut {NoStop}%
\bibitem [{\citenamefont {Bauer}\ \emph {et~al.}(2012)\citenamefont {Bauer}, \citenamefont {Saitoh},\ and\ \citenamefont {van Wees}}]{bauer2012caloritronics}%
  \BibitemOpen
  \bibfield  {author} {\bibinfo {author} {\bibfnamefont {G.~E.~W.}\ \bibnamefont {Bauer}}, \bibinfo {author} {\bibfnamefont {E.}~\bibnamefont {Saitoh}},\ and\ \bibinfo {author} {\bibfnamefont {B.~J.}\ \bibnamefont {van Wees}},\ }\href {https://doi.org/10.1038/nmat3301} {\bibfield  {journal} {\bibinfo  {journal} {Nature Materials}\ }\textbf {\bibinfo {volume} {11}},\ \bibinfo {pages} {391} (\bibinfo {year} {2012})}\BibitemShut {NoStop}%
\bibitem [{\citenamefont {Kuschel}\ \emph {et~al.}(2019)\citenamefont {Kuschel}, \citenamefont {Czerner}, \citenamefont {Walowski}, \citenamefont {Thomas}, \citenamefont {Schumacher}, \citenamefont {Reiss}, \citenamefont {Heiliger},\ and\ \citenamefont {M{\"u}nzenberg}}]{kuschel2019tunnel}%
  \BibitemOpen
  \bibfield  {author} {\bibinfo {author} {\bibfnamefont {T.}~\bibnamefont {Kuschel}}, \bibinfo {author} {\bibfnamefont {M.}~\bibnamefont {Czerner}}, \bibinfo {author} {\bibfnamefont {J.}~\bibnamefont {Walowski}}, \bibinfo {author} {\bibfnamefont {A.}~\bibnamefont {Thomas}}, \bibinfo {author} {\bibfnamefont {H.~W.}\ \bibnamefont {Schumacher}}, \bibinfo {author} {\bibfnamefont {G.}~\bibnamefont {Reiss}}, \bibinfo {author} {\bibfnamefont {C.}~\bibnamefont {Heiliger}},\ and\ \bibinfo {author} {\bibfnamefont {M.}~\bibnamefont {M{\"u}nzenberg}},\ }\href {https://doi.org/10.1088/1361-6463/aafa5f} {\bibfield  {journal} {\bibinfo  {journal} {J. Phys. D: Appl. Phys.}\ }\textbf {\bibinfo {volume} {52}},\ \bibinfo {pages} {133001} (\bibinfo {year} {2019})}\BibitemShut {NoStop}%
\bibitem [{\citenamefont {Walter}\ \emph {et~al.}(2011)\citenamefont {Walter}, \citenamefont {Walowski}, \citenamefont {Zbarsky}, \citenamefont {M{\"u}nzenberg}, \citenamefont {Sch{\"a}fers}, \citenamefont {Ebke}, \citenamefont {Reiss}, \citenamefont {Thomas}, \citenamefont {Peretzki}, \citenamefont {Seibt} \emph {et~al.}}]{walter2011seebeck}%
  \BibitemOpen
  \bibfield  {author} {\bibinfo {author} {\bibfnamefont {M.}~\bibnamefont {Walter}}, \bibinfo {author} {\bibfnamefont {J.}~\bibnamefont {Walowski}}, \bibinfo {author} {\bibfnamefont {V.}~\bibnamefont {Zbarsky}}, \bibinfo {author} {\bibfnamefont {M.}~\bibnamefont {M{\"u}nzenberg}}, \bibinfo {author} {\bibfnamefont {M.}~\bibnamefont {Sch{\"a}fers}}, \bibinfo {author} {\bibfnamefont {D.}~\bibnamefont {Ebke}}, \bibinfo {author} {\bibfnamefont {G.}~\bibnamefont {Reiss}}, \bibinfo {author} {\bibfnamefont {A.}~\bibnamefont {Thomas}}, \bibinfo {author} {\bibfnamefont {P.}~\bibnamefont {Peretzki}}, \bibinfo {author} {\bibfnamefont {M.}~\bibnamefont {Seibt}}, \emph {et~al.},\ }\href {https://doi.org/10.1038/nmat3076} {\bibfield  {journal} {\bibinfo  {journal} {Nat. Mater.}\ }\textbf {\bibinfo {volume} {10}},\ \bibinfo {pages} {742} (\bibinfo {year} {2011})}\BibitemShut {NoStop}%
\bibitem [{\citenamefont {Boehnke}\ \emph {et~al.}(2017)\citenamefont {Boehnke}, \citenamefont {Martens}, \citenamefont {Sterwerf}, \citenamefont {Niesen}, \citenamefont {Huebner}, \citenamefont {von~der Ehe}, \citenamefont {Meinert}, \citenamefont {Kuschel}, \citenamefont {Thomas}, \citenamefont {Heiliger} \emph {et~al.}}]{boehnke2017large}%
  \BibitemOpen
  \bibfield  {author} {\bibinfo {author} {\bibfnamefont {A.}~\bibnamefont {Boehnke}}, \bibinfo {author} {\bibfnamefont {U.}~\bibnamefont {Martens}}, \bibinfo {author} {\bibfnamefont {C.}~\bibnamefont {Sterwerf}}, \bibinfo {author} {\bibfnamefont {A.}~\bibnamefont {Niesen}}, \bibinfo {author} {\bibfnamefont {T.}~\bibnamefont {Huebner}}, \bibinfo {author} {\bibfnamefont {M.}~\bibnamefont {von~der Ehe}}, \bibinfo {author} {\bibfnamefont {M.}~\bibnamefont {Meinert}}, \bibinfo {author} {\bibfnamefont {T.}~\bibnamefont {Kuschel}}, \bibinfo {author} {\bibfnamefont {A.}~\bibnamefont {Thomas}}, \bibinfo {author} {\bibfnamefont {C.}~\bibnamefont {Heiliger}}, \emph {et~al.},\ }\href {https://doi.org/10.1038/s41467-017-01784-x} {\bibfield  {journal} {\bibinfo  {journal} {Nat. Commun.}\ }\textbf {\bibinfo {volume} {8}},\ \bibinfo {pages} {1626} (\bibinfo {year} {2017})}\BibitemShut {NoStop}%
\bibitem [{\citenamefont {Boehnke}\ \emph {et~al.}(2015)\citenamefont {Boehnke}, \citenamefont {Milnikel}, \citenamefont {von~der Ehe}, \citenamefont {Franz}, \citenamefont {Zbarsky}, \citenamefont {Czerner}, \citenamefont {Rott}, \citenamefont {Thomas}, \citenamefont {Heiliger}, \citenamefont {Reiss} \emph {et~al.}}]{boehnke2015off}%
  \BibitemOpen
  \bibfield  {author} {\bibinfo {author} {\bibfnamefont {A.}~\bibnamefont {Boehnke}}, \bibinfo {author} {\bibfnamefont {M.}~\bibnamefont {Milnikel}}, \bibinfo {author} {\bibfnamefont {M.}~\bibnamefont {von~der Ehe}}, \bibinfo {author} {\bibfnamefont {C.}~\bibnamefont {Franz}}, \bibinfo {author} {\bibfnamefont {V.}~\bibnamefont {Zbarsky}}, \bibinfo {author} {\bibfnamefont {M.}~\bibnamefont {Czerner}}, \bibinfo {author} {\bibfnamefont {K.}~\bibnamefont {Rott}}, \bibinfo {author} {\bibfnamefont {A.}~\bibnamefont {Thomas}}, \bibinfo {author} {\bibfnamefont {C.}~\bibnamefont {Heiliger}}, \bibinfo {author} {\bibfnamefont {G.}~\bibnamefont {Reiss}}, \emph {et~al.},\ }\href {https://doi.org/10.1038/srep08945} {\bibfield  {journal} {\bibinfo  {journal} {Sci. Rep.}\ }\textbf {\bibinfo {volume} {5}},\ \bibinfo {pages} {8945} (\bibinfo {year} {2015})}\BibitemShut {NoStop}%
\bibitem [{\citenamefont {Oberbauer}\ \emph {et~al.}(2025{\natexlab{a}})\citenamefont {Oberbauer}, \citenamefont {Winkel}, \citenamefont {Böhnert}, \citenamefont {Claro}, \citenamefont {Benetti}, \citenamefont {Çaha}, \citenamefont {Francis}, \citenamefont {Moradi}, \citenamefont {Ferreira}, \citenamefont {Münzenberg},\ and\ \citenamefont {Parvini}}]{oberbauer2025hybrid}%
  \BibitemOpen
  \bibfield  {author} {\bibinfo {author} {\bibfnamefont {F.}~\bibnamefont {Oberbauer}}, \bibinfo {author} {\bibfnamefont {T.~J.}\ \bibnamefont {Winkel}}, \bibinfo {author} {\bibfnamefont {T.}~\bibnamefont {Böhnert}}, \bibinfo {author} {\bibfnamefont {M.~S.}\ \bibnamefont {Claro}}, \bibinfo {author} {\bibfnamefont {L.}~\bibnamefont {Benetti}}, \bibinfo {author} {\bibfnamefont {I.}~\bibnamefont {Çaha}}, \bibinfo {author} {\bibfnamefont {L.}~\bibnamefont {Francis}}, \bibinfo {author} {\bibfnamefont {F.}~\bibnamefont {Moradi}}, \bibinfo {author} {\bibfnamefont {R.}~\bibnamefont {Ferreira}}, \bibinfo {author} {\bibfnamefont {M.}~\bibnamefont {Münzenberg}},\ and\ \bibinfo {author} {\bibfnamefont {T.~S.}\ \bibnamefont {Parvini}},\ }\href {https://arxiv.org/abs/2501.00813} {\bibinfo {title} {Hybrid opto-electrical excitation of spin-transfer torque nano-oscillators for advanced computing}} (\bibinfo {year} {2025}{\natexlab{a}}),\ \Eprint {https://arxiv.org/abs/2501.00813} {arXiv:2501.00813 [physics.optics]}
  \BibitemShut {NoStop}%
\bibitem [{\citenamefont {Tulapurkar}\ \emph {et~al.}(2005)\citenamefont {Tulapurkar}, \citenamefont {Suzuki}, \citenamefont {Fukushima}, \citenamefont {Kubota}, \citenamefont {Maehara}, \citenamefont {Tsunekawa}, \citenamefont {Djayaprawira}, \citenamefont {Watanabe},\ and\ \citenamefont {Yuasa}}]{tulapurkar2005diode}%
  \BibitemOpen
  \bibfield  {author} {\bibinfo {author} {\bibfnamefont {A.~A.}\ \bibnamefont {Tulapurkar}}, \bibinfo {author} {\bibfnamefont {Y.}~\bibnamefont {Suzuki}}, \bibinfo {author} {\bibfnamefont {A.}~\bibnamefont {Fukushima}}, \bibinfo {author} {\bibfnamefont {H.}~\bibnamefont {Kubota}}, \bibinfo {author} {\bibfnamefont {H.}~\bibnamefont {Maehara}}, \bibinfo {author} {\bibfnamefont {K.}~\bibnamefont {Tsunekawa}}, \bibinfo {author} {\bibfnamefont {D.~D.}\ \bibnamefont {Djayaprawira}}, \bibinfo {author} {\bibfnamefont {N.}~\bibnamefont {Watanabe}},\ and\ \bibinfo {author} {\bibfnamefont {S.}~\bibnamefont {Yuasa}},\ }\href {https://doi.org/10.1038/nature04207} {\bibfield  {journal} {\bibinfo  {journal} {Nature}\ }\textbf {\bibinfo {volume} {438}},\ \bibinfo {pages} {339} (\bibinfo {year} {2005})}\BibitemShut {NoStop}%
\bibitem [{\citenamefont {Jenkins}\ \emph {et~al.}(2016)\citenamefont {Jenkins}, \citenamefont {Lebrun}, \citenamefont {Grimaldi}, \citenamefont {Tsunegi}, \citenamefont {Bortolotti}, \citenamefont {Kubota}, \citenamefont {Yakushiji}, \citenamefont {Fukushima}, \citenamefont {de~Loubens}, \citenamefont {Klein}, \citenamefont {Yuasa},\ and\ \citenamefont {Cros}}]{jenkins2016vortex}%
  \BibitemOpen
  \bibfield  {author} {\bibinfo {author} {\bibfnamefont {A.~S.}\ \bibnamefont {Jenkins}}, \bibinfo {author} {\bibfnamefont {R.}~\bibnamefont {Lebrun}}, \bibinfo {author} {\bibfnamefont {E.}~\bibnamefont {Grimaldi}}, \bibinfo {author} {\bibfnamefont {S.}~\bibnamefont {Tsunegi}}, \bibinfo {author} {\bibfnamefont {P.}~\bibnamefont {Bortolotti}}, \bibinfo {author} {\bibfnamefont {H.}~\bibnamefont {Kubota}}, \bibinfo {author} {\bibfnamefont {K.}~\bibnamefont {Yakushiji}}, \bibinfo {author} {\bibfnamefont {A.}~\bibnamefont {Fukushima}}, \bibinfo {author} {\bibfnamefont {G.}~\bibnamefont {de~Loubens}}, \bibinfo {author} {\bibfnamefont {O.}~\bibnamefont {Klein}}, \bibinfo {author} {\bibfnamefont {S.}~\bibnamefont {Yuasa}},\ and\ \bibinfo {author} {\bibfnamefont {V.}~\bibnamefont {Cros}},\ }\href {https://doi.org/10.1038/nnano.2015.295} {\bibfield  {journal} {\bibinfo  {journal} {Nat. Nanotechnol.}\ }\textbf {\bibinfo {volume} {11}},\ \bibinfo {pages} {360} (\bibinfo {year} {2016})}\BibitemShut {NoStop}%
\bibitem [{\citenamefont {Leroux}\ \emph {et~al.}(2021)\citenamefont {Leroux}, \citenamefont {Mizrahi}, \citenamefont {Markovi{\'c}}, \citenamefont {Sanz-Hern{\'a}ndez}, \citenamefont {Trastoy}, \citenamefont {Bortolotti}, \citenamefont {Martins}, \citenamefont {Jenkins}, \citenamefont {Ferreira},\ and\ \citenamefont {Grollier}}]{leroux2021hardware}%
  \BibitemOpen
  \bibfield  {author} {\bibinfo {author} {\bibfnamefont {N.}~\bibnamefont {Leroux}}, \bibinfo {author} {\bibfnamefont {A.}~\bibnamefont {Mizrahi}}, \bibinfo {author} {\bibfnamefont {D.}~\bibnamefont {Markovi{\'c}}}, \bibinfo {author} {\bibfnamefont {D.}~\bibnamefont {Sanz-Hern{\'a}ndez}}, \bibinfo {author} {\bibfnamefont {J.}~\bibnamefont {Trastoy}}, \bibinfo {author} {\bibfnamefont {P.}~\bibnamefont {Bortolotti}}, \bibinfo {author} {\bibfnamefont {L.}~\bibnamefont {Martins}}, \bibinfo {author} {\bibfnamefont {A.}~\bibnamefont {Jenkins}}, \bibinfo {author} {\bibfnamefont {R.}~\bibnamefont {Ferreira}},\ and\ \bibinfo {author} {\bibfnamefont {J.}~\bibnamefont {Grollier}},\ }\href {https://doi.org/10.1088/2634-4386/abfca6} {\bibfield  {journal} {\bibinfo  {journal} {Neuromorph. Comput. Eng.}\ }\textbf {\bibinfo {volume} {1}},\ \bibinfo {pages} {011001} (\bibinfo {year} {2021})}\BibitemShut {NoStop}%
\bibitem [{\citenamefont {Ross}\ \emph {et~al.}(2023)\citenamefont {Ross}, \citenamefont {Leroux}, \citenamefont {De~Riz}, \citenamefont {Markovi{\'c}}, \citenamefont {Sanz-Hern{\'a}ndez}, \citenamefont {Trastoy}, \citenamefont {Bortolotti}, \citenamefont {Querlioz}, \citenamefont {Martins}, \citenamefont {Benetti}, \citenamefont {Claro}, \citenamefont {Anacleto}, \citenamefont {Schulman}, \citenamefont {Taris}, \citenamefont {Begueret}, \citenamefont {Sa{\"i}ghi}, \citenamefont {Jenkins}, \citenamefont {Ferreira}, \citenamefont {Vincent}, \citenamefont {Mizrahi},\ and\ \citenamefont {Grollier}}]{ross2023multilayer}%
  \BibitemOpen
  \bibfield  {author} {\bibinfo {author} {\bibfnamefont {A.}~\bibnamefont {Ross}}, \bibinfo {author} {\bibfnamefont {N.}~\bibnamefont {Leroux}}, \bibinfo {author} {\bibfnamefont {A.}~\bibnamefont {De~Riz}}, \bibinfo {author} {\bibfnamefont {D.}~\bibnamefont {Markovi{\'c}}}, \bibinfo {author} {\bibfnamefont {D.}~\bibnamefont {Sanz-Hern{\'a}ndez}}, \bibinfo {author} {\bibfnamefont {J.}~\bibnamefont {Trastoy}}, \bibinfo {author} {\bibfnamefont {P.}~\bibnamefont {Bortolotti}}, \bibinfo {author} {\bibfnamefont {D.}~\bibnamefont {Querlioz}}, \bibinfo {author} {\bibfnamefont {L.}~\bibnamefont {Martins}}, \bibinfo {author} {\bibfnamefont {L.}~\bibnamefont {Benetti}}, \bibinfo {author} {\bibfnamefont {M.~S.}\ \bibnamefont {Claro}}, \bibinfo {author} {\bibfnamefont {P.}~\bibnamefont {Anacleto}}, \bibinfo {author} {\bibfnamefont {A.}~\bibnamefont {Schulman}}, \bibinfo {author} {\bibfnamefont {T.}~\bibnamefont {Taris}}, \bibinfo {author} {\bibfnamefont {J.-B.}\ \bibnamefont {Begueret}}, \bibinfo {author} {\bibfnamefont
  {S.}~\bibnamefont {Sa{\"i}ghi}}, \bibinfo {author} {\bibfnamefont {A.~S.}\ \bibnamefont {Jenkins}}, \bibinfo {author} {\bibfnamefont {R.}~\bibnamefont {Ferreira}}, \bibinfo {author} {\bibfnamefont {A.~F.}\ \bibnamefont {Vincent}}, \bibinfo {author} {\bibfnamefont {F.~A.}\ \bibnamefont {Mizrahi}},\ and\ \bibinfo {author} {\bibfnamefont {J.}~\bibnamefont {Grollier}},\ }\href {https://doi.org/10.1038/s41565-023-01452-w} {\bibfield  {journal} {\bibinfo  {journal} {Nat. Nanotechnol.}\ }\textbf {\bibinfo {volume} {18}},\ \bibinfo {pages} {1273} (\bibinfo {year} {2023})}\BibitemShut {NoStop}%
\bibitem [{\citenamefont {Oberbauer}\ \emph {et~al.}(2025{\natexlab{b}})\citenamefont {Oberbauer}, \citenamefont {Winkel}, \citenamefont {B{\"o}hnert}, \citenamefont {Wanjura}, \citenamefont {Claro}, \citenamefont {Benetti}, \citenamefont {{\c C}aha}, \citenamefont {Deepak}, \citenamefont {Moradi}, \citenamefont {Ferreira}, \citenamefont {M{\"u}nzenberg},\ and\ \citenamefont {Parvini}}]{oberbauer2025magnetic}%
  \BibitemOpen
  \bibfield  {author} {\bibinfo {author} {\bibfnamefont {F.}~\bibnamefont {Oberbauer}}, \bibinfo {author} {\bibfnamefont {T.~J.}\ \bibnamefont {Winkel}}, \bibinfo {author} {\bibfnamefont {T.}~\bibnamefont {B{\"o}hnert}}, \bibinfo {author} {\bibfnamefont {C.~C.}\ \bibnamefont {Wanjura}}, \bibinfo {author} {\bibfnamefont {M.~S.}\ \bibnamefont {Claro}}, \bibinfo {author} {\bibfnamefont {L.}~\bibnamefont {Benetti}}, \bibinfo {author} {\bibfnamefont {I.}~\bibnamefont {{\c C}aha}}, \bibinfo {author} {\bibfnamefont {F.~L.}\ \bibnamefont {Deepak}}, \bibinfo {author} {\bibfnamefont {F.}~\bibnamefont {Moradi}}, \bibinfo {author} {\bibfnamefont {R.}~\bibnamefont {Ferreira}}, \bibinfo {author} {\bibfnamefont {M.}~\bibnamefont {M{\"u}nzenberg}},\ and\ \bibinfo {author} {\bibfnamefont {T.~S.}\ \bibnamefont {Parvini}},\ }\href {https://doi.org/10.1038/s42005-025-02257-0} {\bibfield  {journal} {\bibinfo  {journal} {Commun. Phys.}\ }\textbf {\bibinfo {volume} {8}},\ \bibinfo {pages} {329} (\bibinfo {year}
  {2025}{\natexlab{b}})}\BibitemShut {NoStop}%
\bibitem [{\citenamefont {Kirilyuk}\ \emph {et~al.}(2010)\citenamefont {Kirilyuk}, \citenamefont {Kimel},\ and\ \citenamefont {Rasing}}]{kirilyuk2010ultrafast}%
  \BibitemOpen
  \bibfield  {author} {\bibinfo {author} {\bibfnamefont {A.}~\bibnamefont {Kirilyuk}}, \bibinfo {author} {\bibfnamefont {A.~V.}\ \bibnamefont {Kimel}},\ and\ \bibinfo {author} {\bibfnamefont {T.}~\bibnamefont {Rasing}},\ }\href {https://doi.org/10.1103/RevModPhys.82.2731} {\bibfield  {journal} {\bibinfo  {journal} {Rev. Mod. Phys.}\ }\textbf {\bibinfo {volume} {82}},\ \bibinfo {pages} {2731} (\bibinfo {year} {2010})}\BibitemShut {NoStop}%
\bibitem [{\citenamefont {Winkel}\ \emph {et~al.}(2024)\citenamefont {Winkel}, \citenamefont {Parvini}, \citenamefont {Stiewe}, \citenamefont {Walowski}, \citenamefont {Moradi},\ and\ \citenamefont {M{\"u}nzenberg}}]{winkel2024}%
  \BibitemOpen
  \bibfield  {author} {\bibinfo {author} {\bibfnamefont {T.~J.}\ \bibnamefont {Winkel}}, \bibinfo {author} {\bibfnamefont {T.~S.}\ \bibnamefont {Parvini}}, \bibinfo {author} {\bibfnamefont {F.-F.}\ \bibnamefont {Stiewe}}, \bibinfo {author} {\bibfnamefont {J.}~\bibnamefont {Walowski}}, \bibinfo {author} {\bibfnamefont {F.}~\bibnamefont {Moradi}},\ and\ \bibinfo {author} {\bibfnamefont {M.}~\bibnamefont {M{\"u}nzenberg}},\ }\bibfield  {journal} {\bibinfo  {journal} {Appl. Phys. Lett.}\ }\textbf {\bibinfo {volume} {124}},\ \href {https://doi.org/10.1063/5.0183830} {10.1063/5.0183830} (\bibinfo {year} {2024})\BibitemShut {NoStop}%
\bibitem [{\citenamefont {Parvini}\ \emph {et~al.}(2017)\citenamefont {Parvini}, \citenamefont {Tehranchi},\ and\ \citenamefont {Hamidi}}]{parvini2017new}%
  \BibitemOpen
  \bibfield  {author} {\bibinfo {author} {\bibfnamefont {T.~S.}\ \bibnamefont {Parvini}}, \bibinfo {author} {\bibfnamefont {M.}~\bibnamefont {Tehranchi}},\ and\ \bibinfo {author} {\bibfnamefont {S.}~\bibnamefont {Hamidi}},\ }\href {https://doi.org/10.1016/j.photonics.2017.04.003} {\bibfield  {journal} {\bibinfo  {journal} {Photonics Nanostructures: Fundam. Appl.}\ }\textbf {\bibinfo {volume} {25}},\ \bibinfo {pages} {25} (\bibinfo {year} {2017})}\BibitemShut {NoStop}%
\bibitem [{\citenamefont {Parvini}\ and\ \citenamefont {Khazaei~Nezhad}(2022)}]{parvini2022}%
  \BibitemOpen
  \bibfield  {author} {\bibinfo {author} {\bibfnamefont {T.~S.}\ \bibnamefont {Parvini}}\ and\ \bibinfo {author} {\bibfnamefont {M.}~\bibnamefont {Khazaei~Nezhad}},\ }\href {https://doi.org/10.1007/s00340-022-07917-5} {\bibfield  {journal} {\bibinfo  {journal} {Appl. Phys. B}\ }\textbf {\bibinfo {volume} {128}},\ \bibinfo {pages} {194} (\bibinfo {year} {2022})}\BibitemShut {NoStop}%
\bibitem [{\citenamefont {Parvini}\ \emph {et~al.}(2015)\citenamefont {Parvini}, \citenamefont {Tehranchi}, \citenamefont {Hamidi},\ and\ \citenamefont {Sarkarati}}]{parvini2015}%
  \BibitemOpen
  \bibfield  {author} {\bibinfo {author} {\bibfnamefont {T.~S.}\ \bibnamefont {Parvini}}, \bibinfo {author} {\bibfnamefont {M.~M.}\ \bibnamefont {Tehranchi}}, \bibinfo {author} {\bibfnamefont {S.~M.}\ \bibnamefont {Hamidi}},\ and\ \bibinfo {author} {\bibfnamefont {S.}~\bibnamefont {Sarkarati}},\ }\href {https://doi.org/10.1063/1.4934872} {\bibfield  {journal} {\bibinfo  {journal} {J. Appl. Phys.}\ }\textbf {\bibinfo {volume} {118}},\ \bibinfo {pages} {183108} (\bibinfo {year} {2015})}\BibitemShut {NoStop}%
\bibitem [{\citenamefont {Bostr{\"o}m}\ \emph {et~al.}(2021)\citenamefont {Bostr{\"o}m}, \citenamefont {Parvini}, \citenamefont {McIver}, \citenamefont {Rubio}, \citenamefont {Kusminskiy},\ and\ \citenamefont {Sentef}}]{bostrom2021all}%
  \BibitemOpen
  \bibfield  {author} {\bibinfo {author} {\bibfnamefont {E.~V.}\ \bibnamefont {Bostr{\"o}m}}, \bibinfo {author} {\bibfnamefont {T.~S.}\ \bibnamefont {Parvini}}, \bibinfo {author} {\bibfnamefont {J.~W.}\ \bibnamefont {McIver}}, \bibinfo {author} {\bibfnamefont {A.}~\bibnamefont {Rubio}}, \bibinfo {author} {\bibfnamefont {S.~V.}\ \bibnamefont {Kusminskiy}},\ and\ \bibinfo {author} {\bibfnamefont {M.~A.}\ \bibnamefont {Sentef}},\ }\href {https://doi.org/10.1103/PhysRevB.104.L100404} {\bibfield  {journal} {\bibinfo  {journal} {Phys. Rev. B}\ }\textbf {\bibinfo {volume} {104}},\ \bibinfo {pages} {L100404} (\bibinfo {year} {2021})}\BibitemShut {NoStop}%
\bibitem [{\citenamefont {Vi{\~n}as~Bostr{\"o}m}\ \emph {et~al.}(2023)\citenamefont {Vi{\~n}as~Bostr{\"o}m}, \citenamefont {Parvini}, \citenamefont {McIver}, \citenamefont {Rubio}, \citenamefont {Kusminskiy},\ and\ \citenamefont {Sentef}}]{vinas2023direct}%
  \BibitemOpen
  \bibfield  {author} {\bibinfo {author} {\bibfnamefont {E.}~\bibnamefont {Vi{\~n}as~Bostr{\"o}m}}, \bibinfo {author} {\bibfnamefont {T.~S.}\ \bibnamefont {Parvini}}, \bibinfo {author} {\bibfnamefont {J.~W.}\ \bibnamefont {McIver}}, \bibinfo {author} {\bibfnamefont {A.}~\bibnamefont {Rubio}}, \bibinfo {author} {\bibfnamefont {S.~V.}\ \bibnamefont {Kusminskiy}},\ and\ \bibinfo {author} {\bibfnamefont {M.~A.}\ \bibnamefont {Sentef}},\ }\href {https://doi.org/10.1103/PhysRevLett.130.026701} {\bibfield  {journal} {\bibinfo  {journal} {Phys. Rev. Lett.}\ }\textbf {\bibinfo {volume} {130}},\ \bibinfo {pages} {026701} (\bibinfo {year} {2023})}\BibitemShut {NoStop}%
\bibitem [{\citenamefont {Parvini}\ \emph {et~al.}(2025)\citenamefont {Parvini}, \citenamefont {R{\"o}mling}, \citenamefont {Sharma},\ and\ \citenamefont {Kusminskiy}}]{parvini2025}%
  \BibitemOpen
  \bibfield  {author} {\bibinfo {author} {\bibfnamefont {T.~S.}\ \bibnamefont {Parvini}}, \bibinfo {author} {\bibfnamefont {A.-L.~E.}\ \bibnamefont {R{\"o}mling}}, \bibinfo {author} {\bibfnamefont {S.}~\bibnamefont {Sharma}},\ and\ \bibinfo {author} {\bibfnamefont {S.~V.}\ \bibnamefont {Kusminskiy}},\ }\href {https://doi.org/10.1103/PhysRevB.111.014416} {\bibfield  {journal} {\bibinfo  {journal} {Phys. Rev. B}\ }\textbf {\bibinfo {volume} {111}},\ \bibinfo {pages} {014416} (\bibinfo {year} {2025})}\BibitemShut {NoStop}%
\bibitem [{\citenamefont {Parvini}(2026)}]{parvini2026programmable}%
  \BibitemOpen
  \bibfield  {author} {\bibinfo {author} {\bibfnamefont {T.~S.}\ \bibnamefont {Parvini}},\ }\href {https://doi.org/10.1103/qkyh-jpxn} {\bibfield  {journal} {\bibinfo  {journal} {Phys. Rev. B}\ }\textbf {\bibinfo {volume} {113}},\ \bibinfo {pages} {104440} (\bibinfo {year} {2026})}\BibitemShut {NoStop}%
\bibitem [{\citenamefont {Hamidi}\ \emph {et~al.}(2013)\citenamefont {Hamidi}, \citenamefont {Parvini},\ and\ \citenamefont {Tehranchi}}]{hamidi2013}%
  \BibitemOpen
  \bibfield  {author} {\bibinfo {author} {\bibfnamefont {S.}~\bibnamefont {Hamidi}}, \bibinfo {author} {\bibfnamefont {T.~S.}\ \bibnamefont {Parvini}},\ and\ \bibinfo {author} {\bibfnamefont {M.}~\bibnamefont {Tehranchi}},\ }\href {https://doi.org/10.1007/s00339-013-7599-1} {\bibfield  {journal} {\bibinfo  {journal} {Appl. Phys. A}\ }\textbf {\bibinfo {volume} {111}},\ \bibinfo {pages} {525} (\bibinfo {year} {2013})}\BibitemShut {NoStop}%
\bibitem [{\citenamefont {Su}\ \emph {et~al.}(2024)\citenamefont {Su} \emph {et~al.}}]{sukwon2024thermal}%
  \BibitemOpen
  \bibfield  {author} {\bibinfo {author} {\bibfnamefont {K.}~\bibnamefont {Su}} \emph {et~al.},\ }\bibfield  {journal} {\bibinfo  {journal} {Nano Lett.}\ }\href {https://doi.org/10.1021/acs.nanolett.4c02571} {10.1021/acs.nanolett.4c02571} (\bibinfo {year} {2024})\BibitemShut {NoStop}%
\bibitem [{\citenamefont {Jang}\ \emph {et~al.}(2020)\citenamefont {Jang} \emph {et~al.}}]{jang2020thermal}%
  \BibitemOpen
  \bibfield  {author} {\bibinfo {author} {\bibfnamefont {H.}~\bibnamefont {Jang}} \emph {et~al.},\ }\href {https://doi.org/10.1103/PhysRevApplied.13.054006} {\bibfield  {journal} {\bibinfo  {journal} {Phys. Rev. Appl.}\ }\textbf {\bibinfo {volume} {13}},\ \bibinfo {pages} {054006} (\bibinfo {year} {2020})}\BibitemShut {NoStop}%
\bibitem [{\citenamefont {Teixeira}\ \emph {et~al.}(2013)\citenamefont {Teixeira}, \citenamefont {Costa}, \citenamefont {Ventura}, \citenamefont {Fernandez-Garcia}, \citenamefont {Azevedo}, \citenamefont {Araujo}, \citenamefont {Sousa}, \citenamefont {Wisniowski}, \citenamefont {Cardoso},\ and\ \citenamefont {Freitas}}]{teixeira2013giant}%
  \BibitemOpen
  \bibfield  {author} {\bibinfo {author} {\bibfnamefont {J.}~\bibnamefont {Teixeira}}, \bibinfo {author} {\bibfnamefont {J.}~\bibnamefont {Costa}}, \bibinfo {author} {\bibfnamefont {J.}~\bibnamefont {Ventura}}, \bibinfo {author} {\bibfnamefont {M.}~\bibnamefont {Fernandez-Garcia}}, \bibinfo {author} {\bibfnamefont {J.}~\bibnamefont {Azevedo}}, \bibinfo {author} {\bibfnamefont {J.}~\bibnamefont {Araujo}}, \bibinfo {author} {\bibfnamefont {J.}~\bibnamefont {Sousa}}, \bibinfo {author} {\bibfnamefont {P.}~\bibnamefont {Wisniowski}}, \bibinfo {author} {\bibfnamefont {S.}~\bibnamefont {Cardoso}},\ and\ \bibinfo {author} {\bibfnamefont {P.}~\bibnamefont {Freitas}},\ }\bibfield  {journal} {\bibinfo  {journal} {Appl. Phys. Lett.}\ }\textbf {\bibinfo {volume} {102}},\ \href {https://doi.org/10.1063/1.4809569} {10.1063/1.4809569} (\bibinfo {year} {2013})\BibitemShut {NoStop}%
\bibitem [{\citenamefont {Czerner}\ \emph {et~al.}(2011)\citenamefont {Czerner}, \citenamefont {Bachmann},\ and\ \citenamefont {Heiliger}}]{czerner2011spin}%
  \BibitemOpen
  \bibfield  {author} {\bibinfo {author} {\bibfnamefont {M.}~\bibnamefont {Czerner}}, \bibinfo {author} {\bibfnamefont {M.}~\bibnamefont {Bachmann}},\ and\ \bibinfo {author} {\bibfnamefont {C.}~\bibnamefont {Heiliger}},\ }\href {https://doi.org/10.1103/PhysRevB.83.132405} {\bibfield  {journal} {\bibinfo  {journal} {Phys. Rev. B}\ }\textbf {\bibinfo {volume} {83}},\ \bibinfo {pages} {132405} (\bibinfo {year} {2011})}\BibitemShut {NoStop}%
\bibitem [{\citenamefont {Onsager}(1931{\natexlab{a}})}]{onsager1931reciprocalI}%
  \BibitemOpen
  \bibfield  {author} {\bibinfo {author} {\bibfnamefont {L.}~\bibnamefont {Onsager}},\ }\href {https://doi.org/10.1103/PhysRev.37.405} {\bibfield  {journal} {\bibinfo  {journal} {Phys. Rev.}\ }\textbf {\bibinfo {volume} {37}},\ \bibinfo {pages} {405} (\bibinfo {year} {1931}{\natexlab{a}})}\BibitemShut {NoStop}%
\bibitem [{\citenamefont {Onsager}(1931{\natexlab{b}})}]{onsager1931reciprocalII}%
  \BibitemOpen
  \bibfield  {author} {\bibinfo {author} {\bibfnamefont {L.}~\bibnamefont {Onsager}},\ }\href {https://doi.org/10.1103/PhysRev.38.2265} {\bibfield  {journal} {\bibinfo  {journal} {Phys. Rev.}\ }\textbf {\bibinfo {volume} {38}},\ \bibinfo {pages} {2265} (\bibinfo {year} {1931}{\natexlab{b}})}\BibitemShut {NoStop}%
\bibitem [{\citenamefont {Uchida}\ \emph {et~al.}(2008)\citenamefont {Uchida}, \citenamefont {Takahashi}, \citenamefont {Harii}, \citenamefont {Ieda}, \citenamefont {Koshibae}, \citenamefont {Ando}, \citenamefont {Maekawa},\ and\ \citenamefont {Saitoh}}]{uchida2008observation}%
  \BibitemOpen
  \bibfield  {author} {\bibinfo {author} {\bibfnamefont {K.-I.}\ \bibnamefont {Uchida}}, \bibinfo {author} {\bibfnamefont {S.}~\bibnamefont {Takahashi}}, \bibinfo {author} {\bibfnamefont {K.}~\bibnamefont {Harii}}, \bibinfo {author} {\bibfnamefont {J.}~\bibnamefont {Ieda}}, \bibinfo {author} {\bibfnamefont {W.}~\bibnamefont {Koshibae}}, \bibinfo {author} {\bibfnamefont {K.}~\bibnamefont {Ando}}, \bibinfo {author} {\bibfnamefont {S.}~\bibnamefont {Maekawa}},\ and\ \bibinfo {author} {\bibfnamefont {E.}~\bibnamefont {Saitoh}},\ }\href {https://doi.org/10.1038/nature07321} {\bibfield  {journal} {\bibinfo  {journal} {Nature}\ }\textbf {\bibinfo {volume} {455}},\ \bibinfo {pages} {778} (\bibinfo {year} {2008})}\BibitemShut {NoStop}%
\bibitem [{\citenamefont {Gravier}\ \emph {et~al.}(2006)\citenamefont {Gravier}, \citenamefont {Serrano-Guisan}, \citenamefont {Reuse},\ and\ \citenamefont {Ansermet}}]{gravier2006thermodynamic}%
  \BibitemOpen
  \bibfield  {author} {\bibinfo {author} {\bibfnamefont {L.}~\bibnamefont {Gravier}}, \bibinfo {author} {\bibfnamefont {S.}~\bibnamefont {Serrano-Guisan}}, \bibinfo {author} {\bibfnamefont {F.}~\bibnamefont {Reuse}},\ and\ \bibinfo {author} {\bibfnamefont {J.-P.}\ \bibnamefont {Ansermet}},\ }\href {https://doi.org/10.1103/PhysRevB.73.024419} {\bibfield  {journal} {\bibinfo  {journal} {Phys. Rev. B}\ }\textbf {\bibinfo {volume} {73}},\ \bibinfo {pages} {024419} (\bibinfo {year} {2006})}\BibitemShut {NoStop}%
\bibitem [{\citenamefont {Schmidt}\ \emph {et~al.}(2018)\citenamefont {Schmidt}, \citenamefont {Wilken}, \citenamefont {Nunner},\ and\ \citenamefont {Brouwer}}]{schmidt2018boltzmann}%
  \BibitemOpen
  \bibfield  {author} {\bibinfo {author} {\bibfnamefont {R.}~\bibnamefont {Schmidt}}, \bibinfo {author} {\bibfnamefont {F.}~\bibnamefont {Wilken}}, \bibinfo {author} {\bibfnamefont {T.~S.}\ \bibnamefont {Nunner}},\ and\ \bibinfo {author} {\bibfnamefont {P.~W.}\ \bibnamefont {Brouwer}},\ }\href {https://doi.org/10.1103/PhysRevB.98.134421} {\bibfield  {journal} {\bibinfo  {journal} {Phys. Rev. B}\ }\textbf {\bibinfo {volume} {98}},\ \bibinfo {pages} {134421} (\bibinfo {year} {2018})}\BibitemShut {NoStop}%
\bibitem [{\citenamefont {Jha}\ \emph {et~al.}(2023)\citenamefont {Jha}, \citenamefont {Pariyar}, \citenamefont {Parvini}, \citenamefont {Denker}, \citenamefont {Vardhanapu}, \citenamefont {Vijaykumar}, \citenamefont {Ahrens}, \citenamefont {Meyer}, \citenamefont {Seibt}, \citenamefont {Atodiresei} \emph {et~al.}}]{jha2023interface}%
  \BibitemOpen
  \bibfield  {author} {\bibinfo {author} {\bibfnamefont {N.}~\bibnamefont {Jha}}, \bibinfo {author} {\bibfnamefont {A.}~\bibnamefont {Pariyar}}, \bibinfo {author} {\bibfnamefont {T.~S.}\ \bibnamefont {Parvini}}, \bibinfo {author} {\bibfnamefont {C.}~\bibnamefont {Denker}}, \bibinfo {author} {\bibfnamefont {P.~K.}\ \bibnamefont {Vardhanapu}}, \bibinfo {author} {\bibfnamefont {G.}~\bibnamefont {Vijaykumar}}, \bibinfo {author} {\bibfnamefont {A.}~\bibnamefont {Ahrens}}, \bibinfo {author} {\bibfnamefont {T.}~\bibnamefont {Meyer}}, \bibinfo {author} {\bibfnamefont {M.}~\bibnamefont {Seibt}}, \bibinfo {author} {\bibfnamefont {N.}~\bibnamefont {Atodiresei}}, \emph {et~al.},\ }\href {https://doi.org/10.1021/acsaelm.2c01428} {\bibfield  {journal} {\bibinfo  {journal} {ACS Appl. Electron. Mater.}\ }\textbf {\bibinfo {volume} {5}},\ \bibinfo {pages} {1471} (\bibinfo {year} {2023})}\BibitemShut {NoStop}%
\bibitem [{\citenamefont {Parvini}\ \emph {et~al.}(2023)\citenamefont {Parvini}, \citenamefont {Paz}, \citenamefont {B{\"o}hnert}, \citenamefont {Schulman}, \citenamefont {Benetti}, \citenamefont {Oberbauer}, \citenamefont {Walowski}, \citenamefont {Moradi}, \citenamefont {Ferreira},\ and\ \citenamefont {M{\"u}nzenberg}}]{sadat2023enhancing}%
  \BibitemOpen
  \bibfield  {author} {\bibinfo {author} {\bibfnamefont {T.~S.}\ \bibnamefont {Parvini}}, \bibinfo {author} {\bibfnamefont {E.}~\bibnamefont {Paz}}, \bibinfo {author} {\bibfnamefont {T.}~\bibnamefont {B{\"o}hnert}}, \bibinfo {author} {\bibfnamefont {A.}~\bibnamefont {Schulman}}, \bibinfo {author} {\bibfnamefont {L.}~\bibnamefont {Benetti}}, \bibinfo {author} {\bibfnamefont {F.}~\bibnamefont {Oberbauer}}, \bibinfo {author} {\bibfnamefont {J.}~\bibnamefont {Walowski}}, \bibinfo {author} {\bibfnamefont {F.}~\bibnamefont {Moradi}}, \bibinfo {author} {\bibfnamefont {R.}~\bibnamefont {Ferreira}},\ and\ \bibinfo {author} {\bibfnamefont {M.}~\bibnamefont {M{\"u}nzenberg}},\ }\href {https://pubs.aip.org/aip/jap/article-abstract/133/24/243902/2900329/Enhancing-spin-transfer-torque-in-magnetic-tunnel?redirectedFrom=fulltext} {\bibfield  {journal} {\bibinfo  {journal} {J. Appl. Phys.}\ }\textbf {\bibinfo {volume} {133}} (\bibinfo {year} {2023})}\BibitemShut {NoStop}%
\bibitem [{\citenamefont {B{\"o}hnert}\ \emph {et~al.}(2017)\citenamefont {B{\"o}hnert}, \citenamefont {Dutra}, \citenamefont {Sommer}, \citenamefont {Paz}, \citenamefont {Serrano-Guisan}, \citenamefont {Ferreira},\ and\ \citenamefont {Freitas}}]{boehnke2017Bohnert}%
  \BibitemOpen
  \bibfield  {author} {\bibinfo {author} {\bibfnamefont {T.}~\bibnamefont {B{\"o}hnert}}, \bibinfo {author} {\bibfnamefont {R.}~\bibnamefont {Dutra}}, \bibinfo {author} {\bibfnamefont {R.~L.}\ \bibnamefont {Sommer}}, \bibinfo {author} {\bibfnamefont {E.}~\bibnamefont {Paz}}, \bibinfo {author} {\bibfnamefont {S.}~\bibnamefont {Serrano-Guisan}}, \bibinfo {author} {\bibfnamefont {R.}~\bibnamefont {Ferreira}},\ and\ \bibinfo {author} {\bibfnamefont {P.~P.}\ \bibnamefont {Freitas}},\ }\href {https://doi.org/10.1103/PhysRevB.95.104441} {\bibfield  {journal} {\bibinfo  {journal} {Phys. Rev. B}\ }\textbf {\bibinfo {volume} {95}},\ \bibinfo {pages} {104441} (\bibinfo {year} {2017})}\BibitemShut {NoStop}%
\bibitem [{\citenamefont {Bean}\ \emph {et~al.}(2017)\citenamefont {Bean}, \citenamefont {Szyniszewski}, \citenamefont {Chair},\ and\ \citenamefont {McKenna}}]{bean2017atomic}%
  \BibitemOpen
  \bibfield  {author} {\bibinfo {author} {\bibfnamefont {J.~J.}\ \bibnamefont {Bean}}, \bibinfo {author} {\bibfnamefont {M.}~\bibnamefont {Szyniszewski}}, \bibinfo {author} {\bibfnamefont {J.}~\bibnamefont {Chair}},\ and\ \bibinfo {author} {\bibfnamefont {K.~P.}\ \bibnamefont {McKenna}},\ }\href {https://doi.org/10.1038/srep45594} {\bibfield  {journal} {\bibinfo  {journal} {Sci. Rep.}\ }\textbf {\bibinfo {volume} {7}},\ \bibinfo {pages} {45594} (\bibinfo {year} {2017})}\BibitemShut {NoStop}%
\bibitem [{\citenamefont {McKenna}(2018)}]{mckenna2018stability}%
  \BibitemOpen
  \bibfield  {author} {\bibinfo {author} {\bibfnamefont {K.~P.}\ \bibnamefont {McKenna}},\ }\href {https://doi.org/10.1103/PhysRevMaterials.2.125002} {\bibfield  {journal} {\bibinfo  {journal} {Phys. Rev. Materials}\ }\textbf {\bibinfo {volume} {2}},\ \bibinfo {pages} {125002} (\bibinfo {year} {2018})}\BibitemShut {NoStop}%
\bibitem [{\citenamefont {Kuepferling}\ \emph {et~al.}(2015)\citenamefont {Kuepferling}, \citenamefont {Zullino}, \citenamefont {Sola}, \citenamefont {Van~de Wiele}, \citenamefont {Durin}, \citenamefont {Pasquale}, \citenamefont {Rott}, \citenamefont {Reiss},\ and\ \citenamefont {Bertotti}}]{kuepferling2015vortex}%
  \BibitemOpen
  \bibfield  {author} {\bibinfo {author} {\bibfnamefont {M.}~\bibnamefont {Kuepferling}}, \bibinfo {author} {\bibfnamefont {S.}~\bibnamefont {Zullino}}, \bibinfo {author} {\bibfnamefont {A.}~\bibnamefont {Sola}}, \bibinfo {author} {\bibfnamefont {B.}~\bibnamefont {Van~de Wiele}}, \bibinfo {author} {\bibfnamefont {G.}~\bibnamefont {Durin}}, \bibinfo {author} {\bibfnamefont {M.}~\bibnamefont {Pasquale}}, \bibinfo {author} {\bibfnamefont {K.}~\bibnamefont {Rott}}, \bibinfo {author} {\bibfnamefont {G.}~\bibnamefont {Reiss}},\ and\ \bibinfo {author} {\bibfnamefont {G.}~\bibnamefont {Bertotti}},\ }\bibfield  {journal} {\bibinfo  {journal} {J. Appl. Phys.}\ }\textbf {\bibinfo {volume} {117}},\ \href {https://doi.org/10.1063/1.4908142} {10.1063/1.4908142} (\bibinfo {year} {2015})\BibitemShut {NoStop}%
\bibitem [{\citenamefont {Bohn}\ \emph {et~al.}(2018)\citenamefont {Bohn}, \citenamefont {Durin}, \citenamefont {Correa}, \citenamefont {Machado}, \citenamefont {Della~Pace}, \citenamefont {Chesman},\ and\ \citenamefont {Sommer}}]{bohn2018playing}%
  \BibitemOpen
  \bibfield  {author} {\bibinfo {author} {\bibfnamefont {F.}~\bibnamefont {Bohn}}, \bibinfo {author} {\bibfnamefont {G.}~\bibnamefont {Durin}}, \bibinfo {author} {\bibfnamefont {M.~A.}\ \bibnamefont {Correa}}, \bibinfo {author} {\bibfnamefont {N.~R.}\ \bibnamefont {Machado}}, \bibinfo {author} {\bibfnamefont {R.~D.}\ \bibnamefont {Della~Pace}}, \bibinfo {author} {\bibfnamefont {C.}~\bibnamefont {Chesman}},\ and\ \bibinfo {author} {\bibfnamefont {R.~L.}\ \bibnamefont {Sommer}},\ }\href {https://doi.org/10.1038/s41598-018-29576-3} {\bibfield  {journal} {\bibinfo  {journal} {Sci. Rep.}\ }\textbf {\bibinfo {volume} {8}},\ \bibinfo {pages} {11294} (\bibinfo {year} {2018})}\BibitemShut {NoStop}%
\bibitem [{\citenamefont {Yoo}\ \emph {et~al.}(2012)\citenamefont {Yoo}, \citenamefont {Lee},\ and\ \citenamefont {Kim}}]{yoo2012radial}%
  \BibitemOpen
  \bibfield  {author} {\bibinfo {author} {\bibfnamefont {M.-W.}\ \bibnamefont {Yoo}}, \bibinfo {author} {\bibfnamefont {J.}~\bibnamefont {Lee}},\ and\ \bibinfo {author} {\bibfnamefont {S.-K.}\ \bibnamefont {Kim}},\ }\bibfield  {journal} {\bibinfo  {journal} {Appl. Phys. Lett.}\ }\textbf {\bibinfo {volume} {100}},\ \href {https://doi.org/10.1063/1.4705690} {10.1063/1.4705690} (\bibinfo {year} {2012})\BibitemShut {NoStop}%
\bibitem [{\citenamefont {Jung}\ \emph {et~al.}(2022)\citenamefont {Jung}, \citenamefont {Lee}, \citenamefont {Myung}, \citenamefont {Kim}, \citenamefont {Yoon}, \citenamefont {Kwon}, \citenamefont {Ju}, \citenamefont {Kim}, \citenamefont {Yi}, \citenamefont {Han}, \citenamefont {Kwon}, \citenamefont {Seo}, \citenamefont {Lee}, \citenamefont {Koh}, \citenamefont {Lee}, \citenamefont {Song}, \citenamefont {Choi}, \citenamefont {Ham},\ and\ \citenamefont {Kim}}]{jung2022crossbar}%
  \BibitemOpen
  \bibfield  {author} {\bibinfo {author} {\bibfnamefont {S.}~\bibnamefont {Jung}}, \bibinfo {author} {\bibfnamefont {H.}~\bibnamefont {Lee}}, \bibinfo {author} {\bibfnamefont {S.}~\bibnamefont {Myung}}, \bibinfo {author} {\bibfnamefont {H.}~\bibnamefont {Kim}}, \bibinfo {author} {\bibfnamefont {S.~K.}\ \bibnamefont {Yoon}}, \bibinfo {author} {\bibfnamefont {S.-W.}\ \bibnamefont {Kwon}}, \bibinfo {author} {\bibfnamefont {Y.}~\bibnamefont {Ju}}, \bibinfo {author} {\bibfnamefont {M.}~\bibnamefont {Kim}}, \bibinfo {author} {\bibfnamefont {W.}~\bibnamefont {Yi}}, \bibinfo {author} {\bibfnamefont {S.}~\bibnamefont {Han}}, \bibinfo {author} {\bibfnamefont {B.}~\bibnamefont {Kwon}}, \bibinfo {author} {\bibfnamefont {B.~Y.}\ \bibnamefont {Seo}}, \bibinfo {author} {\bibfnamefont {K.}~\bibnamefont {Lee}}, \bibinfo {author} {\bibfnamefont {G.}~\bibnamefont {Koh}}, \bibinfo {author} {\bibfnamefont {K.}~\bibnamefont {Lee}}, \bibinfo {author} {\bibfnamefont {Y.}~\bibnamefont {Song}}, \bibinfo {author} {\bibfnamefont
  {C.}~\bibnamefont {Choi}}, \bibinfo {author} {\bibfnamefont {D.-H.}\ \bibnamefont {Ham}},\ and\ \bibinfo {author} {\bibfnamefont {S.~J.}\ \bibnamefont {Kim}},\ }\href {https://doi.org/10.1038/s41586-021-04196-6} {\bibfield  {journal} {\bibinfo  {journal} {Nature}\ }\textbf {\bibinfo {volume} {601}},\ \bibinfo {pages} {211} (\bibinfo {year} {2022})}\BibitemShut {NoStop}%
\bibitem [{\citenamefont {Alpaydin}\ and\ \citenamefont {Kaynak}(1998)}]{Alpaydin1998Optical}%
  \BibitemOpen
  \bibfield  {author} {\bibinfo {author} {\bibfnamefont {E.}~\bibnamefont {Alpaydin}}\ and\ \bibinfo {author} {\bibfnamefont {C.}~\bibnamefont {Kaynak}},\ }\href {https://doi.org/10.24432/C50P49} {\bibinfo {title} {{Optical Recognition of Handwritten Digits}}},\ \bibinfo {howpublished} {UCI Machine Learning Repository} (\bibinfo {year} {1998})\BibitemShut {NoStop}%
\bibitem [{\citenamefont {Momeni}\ \emph {et~al.}(2024)\citenamefont {Momeni}, \citenamefont {Rahmani}, \citenamefont {Scellier}, \citenamefont {Wright}, \citenamefont {McMahon}, \citenamefont {Wanjura}, \citenamefont {Li}, \citenamefont {Skalli}, \citenamefont {Berloff}, \citenamefont {Onodera} \emph {et~al.}}]{momeni2024training}%
  \BibitemOpen
  \bibfield  {author} {\bibinfo {author} {\bibfnamefont {A.}~\bibnamefont {Momeni}}, \bibinfo {author} {\bibfnamefont {B.}~\bibnamefont {Rahmani}}, \bibinfo {author} {\bibfnamefont {B.}~\bibnamefont {Scellier}}, \bibinfo {author} {\bibfnamefont {L.~G.}\ \bibnamefont {Wright}}, \bibinfo {author} {\bibfnamefont {P.~L.}\ \bibnamefont {McMahon}}, \bibinfo {author} {\bibfnamefont {C.~C.}\ \bibnamefont {Wanjura}}, \bibinfo {author} {\bibfnamefont {Y.}~\bibnamefont {Li}}, \bibinfo {author} {\bibfnamefont {A.}~\bibnamefont {Skalli}}, \bibinfo {author} {\bibfnamefont {N.~G.}\ \bibnamefont {Berloff}}, \bibinfo {author} {\bibfnamefont {T.}~\bibnamefont {Onodera}}, \emph {et~al.},\ }\bibfield  {journal} {\bibinfo  {journal} {arXiv}\ }\href {https://doi.org/10.48550/arXiv.2406.03372} {10.48550/arXiv.2406.03372} (\bibinfo {year} {2024})\BibitemShut {NoStop}%
\bibitem [{\citenamefont {Scellier}\ and\ \citenamefont {Bengio}(2017)}]{scellier2017equilibrium}%
  \BibitemOpen
  \bibfield  {author} {\bibinfo {author} {\bibfnamefont {B.}~\bibnamefont {Scellier}}\ and\ \bibinfo {author} {\bibfnamefont {Y.}~\bibnamefont {Bengio}},\ }\href {https://doi.org/10.3389/fncom.2017.00024} {\bibfield  {journal} {\bibinfo  {journal} {Front. comput. neurosci.}\ }\textbf {\bibinfo {volume} {11}},\ \bibinfo {pages} {24} (\bibinfo {year} {2017})}\BibitemShut {NoStop}%
\bibitem [{\citenamefont {Cin}\ \emph {et~al.}(2025)\citenamefont {Cin}, \citenamefont {Marquardt},\ and\ \citenamefont {Wanjura}}]{cin2025training}%
  \BibitemOpen
  \bibfield  {author} {\bibinfo {author} {\bibfnamefont {N.~D.}\ \bibnamefont {Cin}}, \bibinfo {author} {\bibfnamefont {F.}~\bibnamefont {Marquardt}},\ and\ \bibinfo {author} {\bibfnamefont {C.~C.}\ \bibnamefont {Wanjura}},\ }\href {https://arxiv.org/abs/2508.11750} {\bibinfo {title} {{Training nonlinear optical neural networks with Scattering Backpropagation}}} (\bibinfo {year} {2025}),\ \Eprint {https://arxiv.org/abs/2508.11750} {arXiv:2508.11750 [physics.optics]} \BibitemShut {NoStop}%
\bibitem [{\citenamefont {Vaswani}\ \emph {et~al.}(2017)\citenamefont {Vaswani}, \citenamefont {Shazeer}, \citenamefont {Parmar}, \citenamefont {Uszkoreit}, \citenamefont {Jones}, \citenamefont {Gomez}, \citenamefont {Kaiser},\ and\ \citenamefont {Polosukhin}}]{vaswani2017attention}%
  \BibitemOpen
  \bibfield  {author} {\bibinfo {author} {\bibfnamefont {A.}~\bibnamefont {Vaswani}}, \bibinfo {author} {\bibfnamefont {N.}~\bibnamefont {Shazeer}}, \bibinfo {author} {\bibfnamefont {N.}~\bibnamefont {Parmar}}, \bibinfo {author} {\bibfnamefont {J.}~\bibnamefont {Uszkoreit}}, \bibinfo {author} {\bibfnamefont {L.}~\bibnamefont {Jones}}, \bibinfo {author} {\bibfnamefont {A.~N.}\ \bibnamefont {Gomez}}, \bibinfo {author} {\bibfnamefont {{\L}.}~\bibnamefont {Kaiser}},\ and\ \bibinfo {author} {\bibfnamefont {I.}~\bibnamefont {Polosukhin}},\ }in\ \href@noop {} {\emph {\bibinfo {booktitle} {Advances in Neural Information Processing Systems}}},\ Vol.~\bibinfo {volume} {30}\ (\bibinfo {year} {2017})\BibitemShut {NoStop}%
\bibitem [{\citenamefont {Hochreiter}\ and\ \citenamefont {Schmidhuber}(1997)}]{hochreiter1997lstm}%
  \BibitemOpen
  \bibfield  {author} {\bibinfo {author} {\bibfnamefont {S.}~\bibnamefont {Hochreiter}}\ and\ \bibinfo {author} {\bibfnamefont {J.}~\bibnamefont {Schmidhuber}},\ }\href {https://doi.org/10.1162/neco.1997.9.8.1735} {\bibfield  {journal} {\bibinfo  {journal} {Neural Computation}\ }\textbf {\bibinfo {volume} {9}},\ \bibinfo {pages} {1735} (\bibinfo {year} {1997})}\BibitemShut {NoStop}%
\bibitem [{\citenamefont {Kipf}\ and\ \citenamefont {Welling}(2017)}]{kipf2017gcn}%
  \BibitemOpen
  \bibfield  {author} {\bibinfo {author} {\bibfnamefont {T.~N.}\ \bibnamefont {Kipf}}\ and\ \bibinfo {author} {\bibfnamefont {M.}~\bibnamefont {Welling}},\ }in\ \href@noop {} {\emph {\bibinfo {booktitle} {International Conference on Learning Representations}}}\ (\bibinfo {year} {2017})\BibitemShut {NoStop}%
\end{thebibliography}%

\end{document}